# Most Strong Lensing Deflectors in the AGEL Survey Are in Group and Cluster Environments

William J. Gottemoller[1,2], Nandini Sahu[1,3], Rodrigo Cordova Rosado[1], Leena Iwamoto[1,2], Courtney B. Watson[1], Kim-Vy H. Tran[1,4], A. Makai Baker[5,6], Tania M. Barone[4,7], Duncan J. Bowden[1,8], Karl Glazebrook[4,7], Anishya Harshan[9], Tucker Jones[10], Glenn G. Kacprzak[4,7], and Camryn M. Neches[1,2]

[1] Center for Astrophysics | Harvard & Smithsonian, 60 Garden Street, Cambridge, MA 02138, USA; williamgottemoller@gmail.com
[2] Department of Astronomy, Harvard University, Cambridge, MA 02138, USA
[3] University of New South Wales, Sydney, NSW 2052, Australia
[4] The ARC Centre of Excellence for All Sky Astrophysics in 3 Dimensions (ASTRO 3D), Australia
[5] School of Physics and Astronomy, Monash University, Clayton VIC 3800, Australia
[6] Department of Physics, University of California, Berkeley, CA 94720, USA
[7] Centre for Astrophysics and Supercomputing, Swinburne University of Technology, PO Box 218, Hawthorn, VIC 3122, Australia
[8] School of Physics & Astronomy, University of Southampton, Southampton SO17 1BJ, UK
[9] University of Ljubljana, Department of Mathematics and Physics, Jadranska ulica 19, SI-1000 Ljubljana, Slovenia
[10] Department of Physics and Astronomy, University of California, Davis, 1 Shields Avenue, Davis, CA 95616, USA

## Abstract

The environments of deflectors in strong-lensing systems affect our ability to test cosmological models and constrain evolutionary properties of galaxies. Here, we measure the deflector scale (Einstein mass) and deflector environment (halo mass) of 89 spectroscopically confirmed strong lenses in the ASTRO3D Galaxy Evolution With Lenses (AGEL) survey. We classify the deflector scale by measuring $\theta_{\rm E}$ to determine the mass enclosed by the Einstein radius, $M(<\theta_{\rm E})$. We quantify deflector environment by using photometric redshifts to determine the galaxy surface density to the fifth-nearest neighbor $\Sigma_5(z)$. We find that 47.2% of our deflectors are embedded in cluster environments, whereas only 9.0% have cluster-scale Einstein radii (masses). We measure a weak correlation ($r = 0.38$) between Einstein mass and $\Sigma_5(z)$, suggesting that the assumption of single galaxy-scale deflectors in lens modeling is overly simplified. We hypothesize that the weak correlation results from galaxy-scale bias in the original AGEL selection and the observational challenge of detecting faint arcs with large Einstein radii. Comparing number densities, $N_{\rm gal}$, between AGEL and control fields, we find that AGEL deflectors are in systematically denser environments. Our study provides a method to identify strong lenses as a function of deflector environment and approximate the impact of large-scale environment in lens modeling. We provide the measured lensing parameters for our 89 AGEL systems as well as $z_{\rm phot}$ and $r$-mag (AB) maps of the line of sight.[11]

## 1. Introduction

Strong gravitational lensing is a powerful tool for exploring a range of astrophysical phenomena and gaining new insight on galaxy evolution and cosmology. These applications include measuring cosmological parameters (e.g., T. E. Collett & M. W. Auger 2014; K. C. Wong et al. 2019; S. Birrer et al. 2022, 2024; N. Sahu et al. 2025; D. J. Bowden et al. 2025), mass profiles of early-type galaxies (e.g., T. Treu & L. V. E. Koopmans 2004; M. Limousin et al. 2005; J. W. Nightingale et al. 2019; A. Sonnenfeld & M. Cautun 2021; N. Sahu et al. 2024), dark matter halo and subhalo characteristics (e.g., Y. Hezaveh et al. 2016; A. J. Shajib et al. 2021; M. Limousin et al. 2022; S. Vegetti et al. 2023; D. Kong et al. 2024), properties of high-redshift sources (e.g., K. G. C. Vasan et al. 2023; A. Ferrara 2024; G. C. K. Vasan et al. 2025), and the circumgalactic medium of quiescent galaxies (e.g., T. M. Barone et al. 2024).

As next-generation surveys—such as those undertaken by Euclid (Euclid Collaboration et al. 2025) and the Vera Rubin Observatory's Legacy Survey of Space and Time (LSST; LSST Science Collaboration et al. 2009)— come online, we will see an explosion in our current lens sample. T. E. Collett (2015) predicts that Euclid and LSST could each detect $\sim 10^5$ galaxy-scale lenses. At larger scales, K. T. Abe et al. (2025) predicts that as many as $\sim 10^2$ quasars and supernovae (SNe) lensed by cluster-scale deflectors could be detected in LSST. The power of so many detections should improve lensing-constrained cosmological and galaxy evolution parameters, including a potential late-Universe sub-1% measurement of $H_0$ (S. Birrer et al. 2024). Batch modeling of lenses could provide crucial, independent confirmation of ΛCDM predictions.

In order to access the constraining power of the new lens samples, future batch lens modeling must account for systematic effects. Several systematic issues may affect lens models used to obtain precise measurements of cosmological and galaxy evolution parameters. Systematics can arise from (1) irregularities in deflector mass models (A. Etherington et al. 2024), (2) the mass-sheet degeneracy (E. E. Falco et al. 1985), or, crucially, (3) the environment of the deflecting mass (C. R. Keeton & A. I. Zabludoff 2004; K. C. Wong et al. 2011;

[11] These resources will be made publicly available at the Harvard Dataverse: doi: 10.7910/DVN/GYDQQP.

M. Jaroszynski & Z. Kostrzewa-Rutkowska 2014; C. R. Keeton et al. 2016).

Strong lens deflectors are classified according to their deflecting mass and host halo masses. The deflecting mass can be grouped into three broad classifications (galaxy-, group-, and cluster-scale deflectors) depending on their measured Einstein mass, $M(<\theta_E)$ (the mass enclosed by the Einstein radius, $\theta_E$). In our work, we refer to the deflector classification using our method as the "deflector scale." Conversely, deflectors can be grouped into three classifications (also galaxy-, group-, and cluster-scale environments) based on the size of their host halo. In our work, we refer to this classification as "deflector environment." As the Einstein mass typically encloses a fraction of the deflector's halo mass, the *deflector scale* and *deflector environment* are not equivalent measurements (M. Limousin et al. 2009; C. Lemon et al. 2024). The disconnect between the deflector scale and deflector environment is a central systematic issue in lens modeling (S. H. Suyu 2012; C. R. Keeton et al. 2016).

Robust modeling of lenses at large deflector scales—particularly cluster-scale lenses—is critical to making independent measurements of cosmological parameters (J. Gilmore & P. Natarajan 2009; G. B. Caminha et al. 2016, 2022; I. Zubeldia & A. Challinor 2019). At these scales, the complex mass distributions of the deflector lead to discrepancies in the lensing potential (P. Natarajan et al. 2024). Some mass models of cluster-scale lenses have reproduced models with fitted source positions $\Delta_{\rm rms} \lesssim 0\overset{''}{.}2$ (G. B. Caminha et al. 2019, 2022; G. V. Pignataro et al. 2021; G. Mahler et al. 2023). However, improving the precision of cosmological measurements requires sub-$0\overset{''}{.}1$ rms (P. Natarajan et al. 2024). Cluster-scale lenses are expected to be comparatively uncommon in near-future lens surveys, limiting their versatility in cosmological studies (K. T. Abe et al. 2025).

Group-scale lenses, on the other hand, are predicted to be $\sim$10 times more abundant than cluster-scale lenses in future surveys (K. T. Abe et al. 2025). Although group-scale lenses have simpler mass distributions, they also have the power to verify $\Lambda$CDM cosmological predictions (A. B. Newman et al. 2015). Thus, group-scale lensing could help complete the continuum between galaxies and clusters in large-scale structure evolution models. Much attention has been given to galaxy- and cluster-scale lenses, but only a few group-scale lenses have been reported (M. Limousin et al. 2009; A. More et al. 2012; A. B. Newman et al. 2015).

Therefore, three problems are central to our analysis. First, systematics play an important role in lens modeling. Second, the discrepancies between the deflector scale and environment are crucial systematic issues. And third, few group-scale lenses have been reported.

To address these three problems, we develop a prescription to simultaneously measure the deflector scale and environment. We do this by mapping deflector environment using photometric redshifts, $z_{\rm phot}$, and deflector scale by measuring Einstein radii, $\theta_E$, using observational data. In Section 2, we discuss the datasets we obtain from DECaLS DR10 (A. Dey et al. 2019) for 89 lenses in the ASTRO3D Galaxy Evolution With Lenses Survey (AGEL DR2; T. M. Barone et al. 2026). In Section 3, we describe how we use Easy and Accurate $z_{\rm phot}$ from Yale (EAZY; G. B. Brammer et al. 2008) to measure photometric redshifts from DECaLS photometry (Section 3.1). We show how we classify the deflector environment density in Section 3.2. We discuss our measurements of the deflector scale using $M(<\theta_E)$ in Section 3.3. We share our deflector scale and deflector environment distributions in Section 4. We consider the relation between the deflector scale and environment, how AGEL fields compare to non-AGEL fields, and takeaways for future lensing analyses in Section 5. Finally, we summarize our conclusions in Section 6.

In our paper, we assume a flat $\Lambda$CDM cosmology with $H_0 = 70\,{\rm km\,s^{-1}\,Mpc^{-1}}$, $\Omega_m = 0.3$, and $\Omega_\Lambda = 0.7$. All magnitudes are expressed in the AB system. All masses are reported in $\log_{10} M_\odot$.

## 2. Observations and Selection

### 2.1. AGEL Survey Dataset

The AGEL sample provides a large repository of lenses at a range of redshifts, deflector scales, and environments. The AGEL[12] survey uses convolutional neural networks (CNNs) from C. Jacobs et al. (2019a) and C. Jacobs et al. (2019b) to detect strong lens systems in the Dark Energy Survey (DES), the Dark Energy Camera Legacy Survey (DECaLS), and Dark Energy Spectroscopic Instrument (DESI) data (K.-V. Tran et al. 2022). To train the CNN, C. Jacobs et al. (2019a) generate a set of 250,000 simulated images of lenses and nonlenses. The CNN is trained to find systems with blue sources near red (early-type) deflectors, while also distinguishing lenses from nonlenses (i.e., ring or spiral galaxies). In K.-V. Tran et al. (2022), the trained CNN is applied to DECaLS DR7 and DES Year 3 red galaxies selected based on magnitude and color cuts. The combined C. Jacobs et al. (2019a) and K.-V. Tran et al. (2022) efforts have identified $\sim 10^3$ lens candidates.

In the second data release (DR2; T. M. Barone et al. 2026), the AGEL team obtained spectroscopic confirmation for 138 systems. The AGEL survey has, so far, detected $\mathcal{O}(10^3)$ strong lens candidates with $r_{\rm mag,\ defl} < 22$ (C. Jacobs et al. 2019a; C. Jacobs et al. 2019b). AGEL DR2 also possesses high-resolution Hubble Space Telescope (HST) Wide-Field Camera 3 (WFC3) imaging in infrared (F140W) and optical (F200LP) bands for 71 systems.

Measurements of the deflector scale and environment require reliable measurements of both the deflector and source redshifts. Accordingly, we consider only those systems with confirmed spectroscopic redshifts of both the source and main deflector. In most cases, we have $z_{\rm spec}$ only for the central (brightest) deflector. We require $z_{\rm defl}$ between 0.2 and 1 to ensure that our $z_{\rm phot}$ measurements precisely constrain deflector environment candidates.

AGEL's spectroscopic sample presents several notable selection biases. For example, observer science preferences may select galaxy-scale lenses for simpler mass modeling or cluster-scale lenses for measurements of the interstellar medium. Spectroscopic observations may also be limited by source separation from the deflector or whether the deflector and source both fit in the integral field unit field. AGEL spectroscopic biases are highly observer dependent and thus difficult to quantify.

The AGEL parent catalog consists only of the lens candidates directly selected by the CNN. Accordingly, it does not possess the same observer-centric and observation-centric biases. The parent catalog also lacks spectroscopic confirmation of deflector and source redshifts. However, because (1) we cannot reliably measure source photometric redshifts and (2) computational limitations, we cannot perform our analysis with the AGEL parent catalog. The reliability and degree of bias in human strong lens selections is a subject of ongoing research, and we refer the reader to K. Rojas et al. (2023) for further information on this subject.

[12] sites.google.com/view/agelsurvey/home?authuser=0

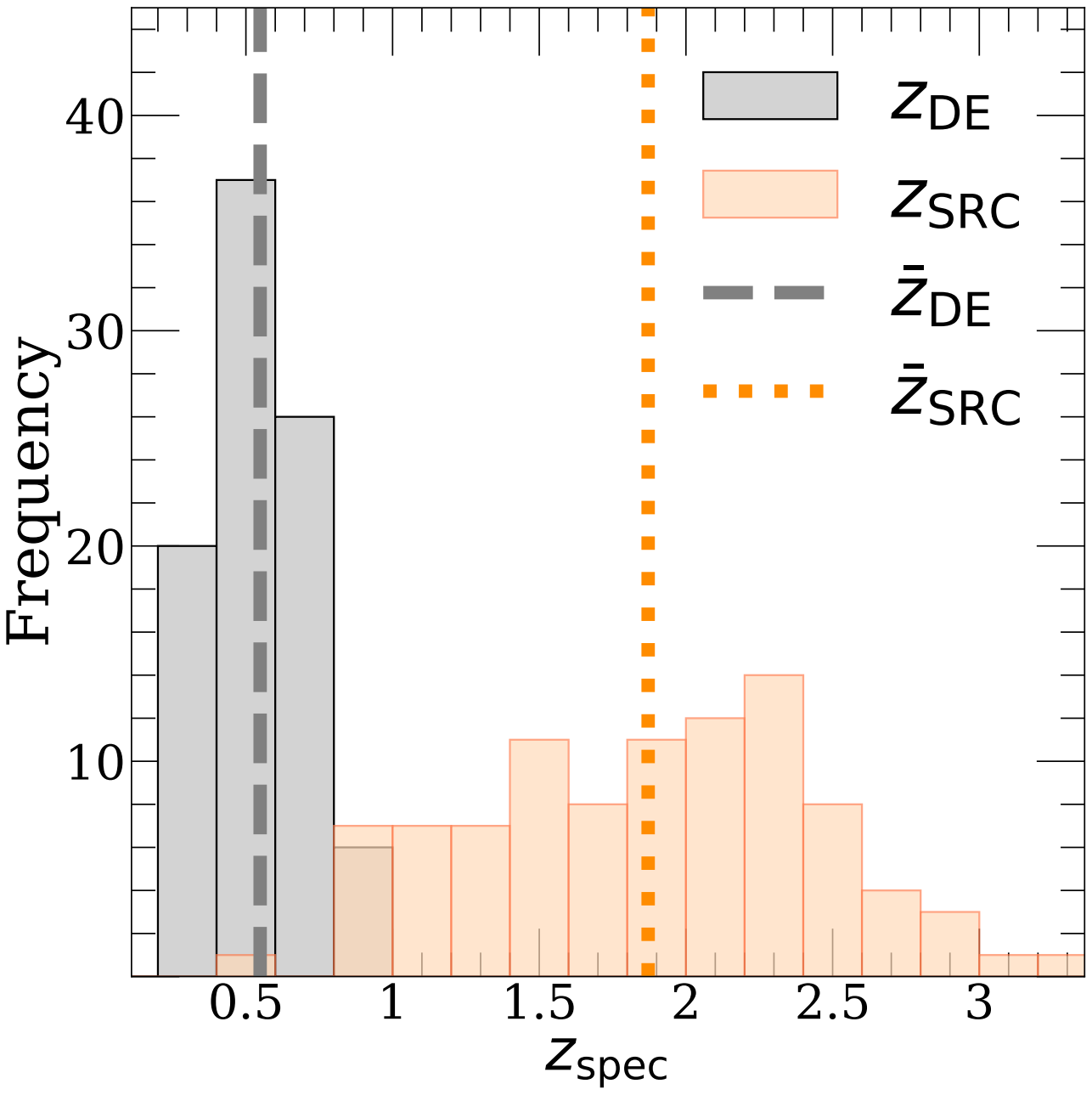

**Figure 1.** The $z_{\rm spec}$ distribution of AGEL deflectors (in gray) and sources (in orange) with DECaLS photometry and spectroscopic $z_{\rm defl}$, $z_{\rm src}$ (89 systems). Our sample has a mean deflector redshift of $\bar{z}_{\rm defl}$ = 0.55 (dashed gray) and source redshift of $\bar{z}_{\rm src}$ = 1.87 (dotted orange).

Choosing only systems with existing photometric data in DECaLS DR10, we obtain DECaLS catalogs for 89 strong lenses. Figure 1 shows the spectroscopic redshift distribution of the deflectors (in orange) and sources (in gray) for our sample. The average $z_{\rm defl}$ and $z_{\rm src}$ are 0.55 and 1.87, respectively.

### 2.2. Selection Methods

For each system, we use TOPCAT (M. B. Taylor 2005) to obtain DECaLS photometry across the $g$, $r$, $z$, and—if available—$i$ bands. We also obtain W1 and W2 mid-IR photometry from the unWISE release of Wide-Field Infrared Survey Explorer data (WISE; E. L. Wright et al. 2010; E. F. Schlafly et al. 2019). We retrieve the list of objects within a 250 kpc radius aperture (corresponding to the proper distance at $z_{\rm defl}$) centered on each central deflector. We choose physical coordinates to account for the wide $z_{\rm defl}$ distribution of our sample (see Figure 1).

DECaLS uses the program Tractor to fit light profiles to detected sources (D. Lang et al. 2016). Even in dense environments, Tractor can reliably measure photometry for many objects (D. Lang et al. 2016; K. Nyland et al. 2017). With that said, the large point-spread function (PSF) for DECaLS and WISE photometry ($1\farcs3$ and $6\farcs1$ in the DECaLS $r$ band and WISE W1 bands, respectively) may cause blending of deflector and source photometry, especially for deflectors with small Einstein radii. We refer the reader to Section 3.1.3 for a discussion on how blending may impact our $z_{\rm phot}$ measurements.

In DECaLS DR10, the $r$ band achieves a 5$\sigma$ detection depth of mag$_r$ = 23.9 (A. Dey et al. 2019), which we adopt as our magnitude limit. The $r$-band point-source depth may not account for low-signal-to-noise ratio (SNR) extended sources, so we apply an $r$-band SNR cut of SNR$_r \geqslant 5$. Although our choice of $r$ band for signal cutoffs is consistent with the prescription in C. Jacobs et al. (2019a), the 4000 Å break will be redshifted out of the $r$ band's bandpass at $z \approx 0.7$. Accordingly, our SNR cutoff may be biased against high-redshift photometry of early-type galaxies. $i$-band photometry lacks that bias, but we choose the $r$ band because $i$-band photometry is incomplete for our systems. For our analysis, we choose all objects that are not point sources ("PSF") or duplicate objects.

To account for incompleteness, we apply an absolute magnitude cut, defined as $+2\sigma$ from the median $r$-band absolute magnitude of all objects with measured $z_{\rm phot} \geqslant 0.8$. This corresponds to an $r$-band absolute magnitude of $-17.62$. Our measurement does not fully account for incompleteness, so we apply additional adjustments in Section 3.2.

## 3. Methodology

### 3.1. Photometric Redshift Measurements

To map the deflector environments of each system, we measure photometric redshifts of all detected objects along the line of sight (LOS). We use the spectral energy distribution (SED) fitting code EAZY (G. B. Brammer et al. 2008) to estimate $z_{\rm phot}$. Most of our deflectors are early-type galaxies, so we exploit the position of the 4000 Å break to constrain the photometric redshifts for our sample. Our measurements provide close $z_{\rm phot}$ constraints on object redshifts.

#### 3.1.1. EAZY Priors

We use the EAZY v1.3 template set and EAZY's standard template error function. Beyond $z \approx 1.5$, the 4000 Å break is outside the transmission of the $z$ band, DECaLS's reddest band. The DECaLS observational data also suffer from increased incompleteness at higher redshifts. For these reasons, we institute a $z_{\rm phot}$ prior of $0.01 \leqslant z_{\rm phot} \leqslant 1.5$.

We use the DECaLS $r$ band as our prior filter. For central deflectors with known $z_{\rm spec}$, we leave $z_{\rm phot}$ as a free parameter to measure the spread in our $z_{\rm phot}$ measurements. We require measurements in at least four bands for $z_{\rm phot}$ estimates. All other parameters are set to their defaults (see zphot.param.default in G. B. Brammer et al. 2008 for more information on default EAZY parameters).

#### 3.1.2. SED Fitting

EAZY outputs best-fit SEDs and the corresponding $z_{\rm phot}$ probability distribution function (PDF) for each fitted object. Figure 2 shows an EAZY output SED fit (left) and $z_{\rm phot}$ $P(z)$ (right) for a galaxy at $z_{\rm phot}$ = 0.80. EAZY effectively constrain the redshifts of many bright early-type galaxies by exploiting the 4000 Å break.

Our SNR and magnitude cutoffs (Section 2.2) remove most poor-signal objects. However, some objects with high $r$-band fluxes may provide poorly constrained $z_{\rm phot}$ measurements. This may be because of a combination of galaxy type (i.e., emission line galaxies), redshift, and blending. To remove poor fits, we apply the "risk" parameter, $R(z_{\rm phot})$ in M. Tanaka et al. (2018). $R(z_{\rm phot})$ quantifies the $z_{\rm phot}$ fit quality from the $P(z)$. It is defined as (M. Tanaka et al. 2018)

$$R(z_{\rm phot}) = \int P(z)\, L\left(\frac{z_{\rm phot} - z}{1 + z}\right) dz, \tag{1}$$

with $L\left(\frac{z_{\rm phot} - z}{1 + z}\right)$, the loss function, defined as

$$L\left(\frac{z_{\rm phot} - z}{1 + z}\right) = 1 - \frac{1}{1 + \left(\frac{\Delta z}{\gamma}\right)^2}, \tag{2}$$

where $\Delta z$ is the difference between $z$ and the fitted $z_{\rm phot}$. In our work, we choose $\gamma = 0.15$—the standard $\Delta z$ at which a $z_{\rm phot}$ fit is classified as an "outlier" (M. Tanaka et al. 2018). Based on a visual inspection of our $z_{\rm phot}$ PDFs, we cut fits with $R(z) \geqslant 0.20$.

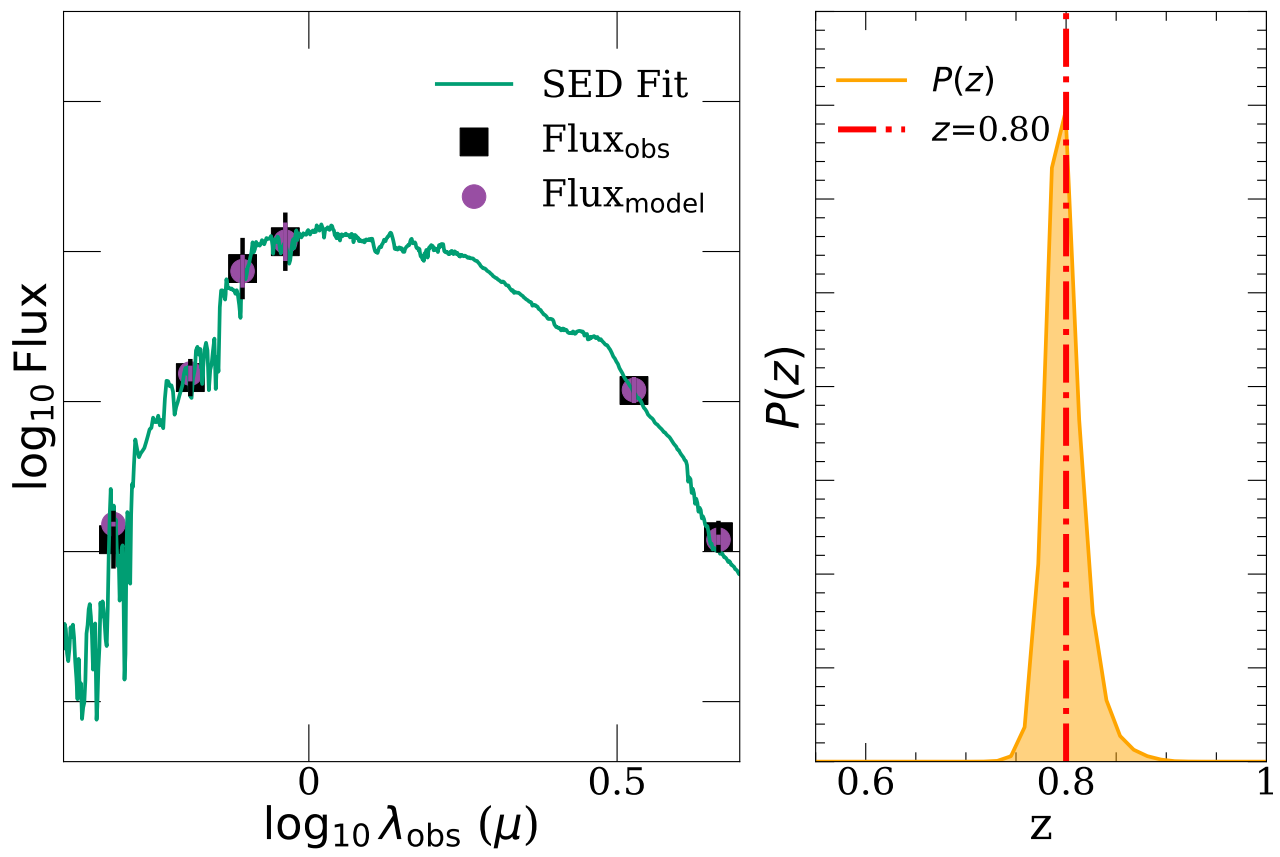

**Figure 2.** EAZY output for a line-of-sight object in the field of AGEL2303, with fitted redshift $z_{\rm phot}$ = 0.80. Left: log$_{10}$ flux versus log$_{10}$ observed wavelength. The black squares are the observed fluxes in six bands, and the magenta circles are the modeled flux measurements; the green distribution is the SED fit. The posterior SED shows continuum features, consistent with an early-type galaxy. Right: the posterior redshift probability distribution, $P(z)$, against $z$.

#### 3.1.3. Accuracy of $z_{phot}$ Measurements

To compare our $z_{\rm phot}$ measurements to spectroscopic ("true") redshifts, we compare $z_{\rm phot}$ and $z_{\rm spec}$ of all our deflectors. Figure 3 shows our $z_{\rm phot}$ measurements against $z_{\rm spec}$ for 83 deflectors; the remaining 6 deflectors were removed by our $R(z)$ cutoff. For SED fits not showing obvious blending of the deflector with source photometry, we measure $\frac{\Delta z}{1 + z_{\rm spec}} = 0.033$ (where $\Delta z = |z_{\rm phot} - z_{\rm spec}|$). Our $z_{\rm phot}$ measurements accurately reproduce $z_{\rm spec}$ for most of our sample.

However, some of our photometry shows blending between deflector and source light. 17 of our deflectors show (1) the presence of emission line features in the source SED and (2) the appearance of blending in image cutouts. The emission line features and unexpected ratios between bandpasses lead to the dilution of the 4000 Å break and the appearance of star-forming galactic SED characteristics, such as emission lines or the Lyman break. We measure a mean $\frac{\Delta z}{1 + z_{\rm spec}} = 0.07$ for our blended deflectors, larger than that measured for unblended deflectors. All but one of our deflectors overestimate the spectroscopic $z_{\rm defl}$.

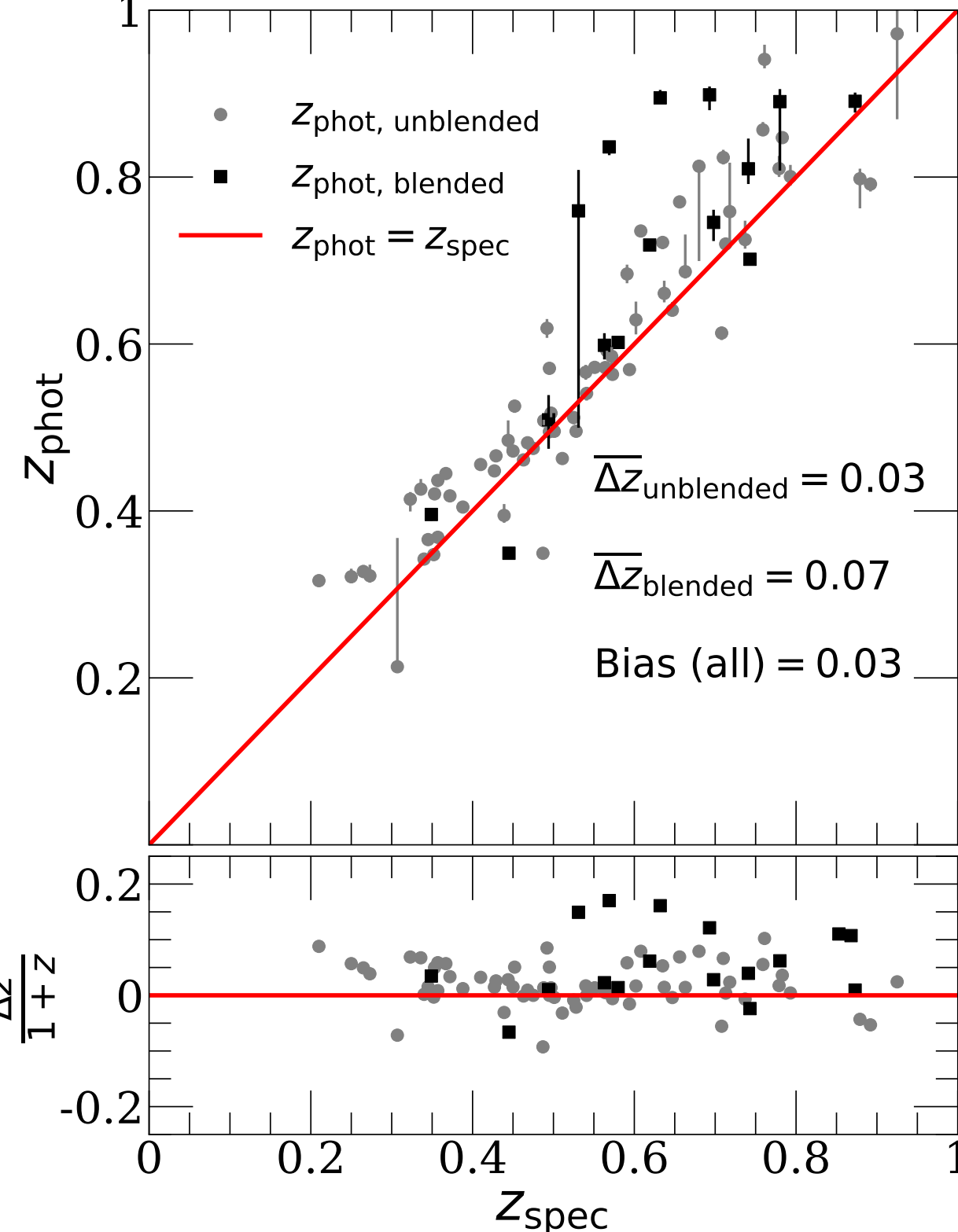

**Figure 3.** $z_{\rm phot}$ versus $z_{\rm spec}$ plot for a sample of 83 central deflectors with measured $z_{\rm phot}$. Black triangles are deflectors whose SED fits showed blending of photometry with the source, and gray circles are deflectors without obvious signs of blending. Error bars are the 16th and 84th percentile values in the $P(z)$ outputs. For unblended systems, we measure $\frac{\Delta z}{1 + z_{\rm spec}} = 0.032$, demonstrating that $z_{\rm phot}$ measurements accurately reproduce deflector redshifts for most AGEL systems. We observe a bias of 0.03 for these fits. For blended systems, we measure $\frac{\Delta z}{1 + z_{\rm spec}}$ of 0.07. The bias in our unblended deflectors is due to $z_{\rm phot}$ overestimates of early-type galaxies (F. B. Abdalla et al. 2011); in our blended deflectors, it is due to source light bleeding into deflector photometry.

Our measurements of unblended deflectors are biased toward higher redshifts (mean bias, 0.03). $z_{\rm phot}$ measurements are known to overestimate the redshift of early-type galaxies (see Figures 4 and 5 in F. B. Abdalla et al. 2011) at lower redshifts, so our bias is not unexpected. In most cases, the $z_{\rm phot}$ cutoffs we apply to identify group candidates are wide enough to account for both the bias and scatter in our measurements.

For objects without $z_{\rm spec}$ measurements, we compare our $z_{\rm phot}$ results to $z_{\rm phot}$ measurements in R. Zhou et al. (2021). We find, over a redshift range of $0 < z < 1$, $\frac{\Delta z}{1 + z} = 0.061$. The larger scatter may be due to discrepancies between our and the R. Zhou et al. (2021) measurement methods. The lower risk of source blending may improve $z_{\rm phot}$ measurements of other LOS objects relative to central deflectors. The bias we observe ($-0.037$) may be due to biases in the low-$z$ sample from R. Zhou et al. (2021).

### 3.2. Deflector Environment Measurements

For each system, we map galaxies along the LOS according to their redshift relative to the deflector, $z_{\rm defl}$. We define "deflector environment candidates" as objects with fitted $z_{\rm phot}$ within 20% of $z_{\rm defl}$. This cutoff is approximately the median width of our $1\sigma$ confidence interval for an object at $z_{\mathrm{phot}} = 0.5$. We classify objects with $z_{\mathrm{phot}} < 0.8 z_{\mathrm{defl}}$ as foreground galaxies, and objects with $z_{\mathrm{phot}} > 1.2 z_{\mathrm{defl}}$ as background objects. Figure 4 shows the LOS of three systems, AGEL0003, AGEL0919, and AGEL0424. The three systems are arranged from left to right by number of deflector environment candidates, $N_{\mathrm{env}}(<250\,\mathrm{kpc})$. The blue squares are foreground candidates, the green circles are deflector environment candidates, and the red triangles are background candidates.

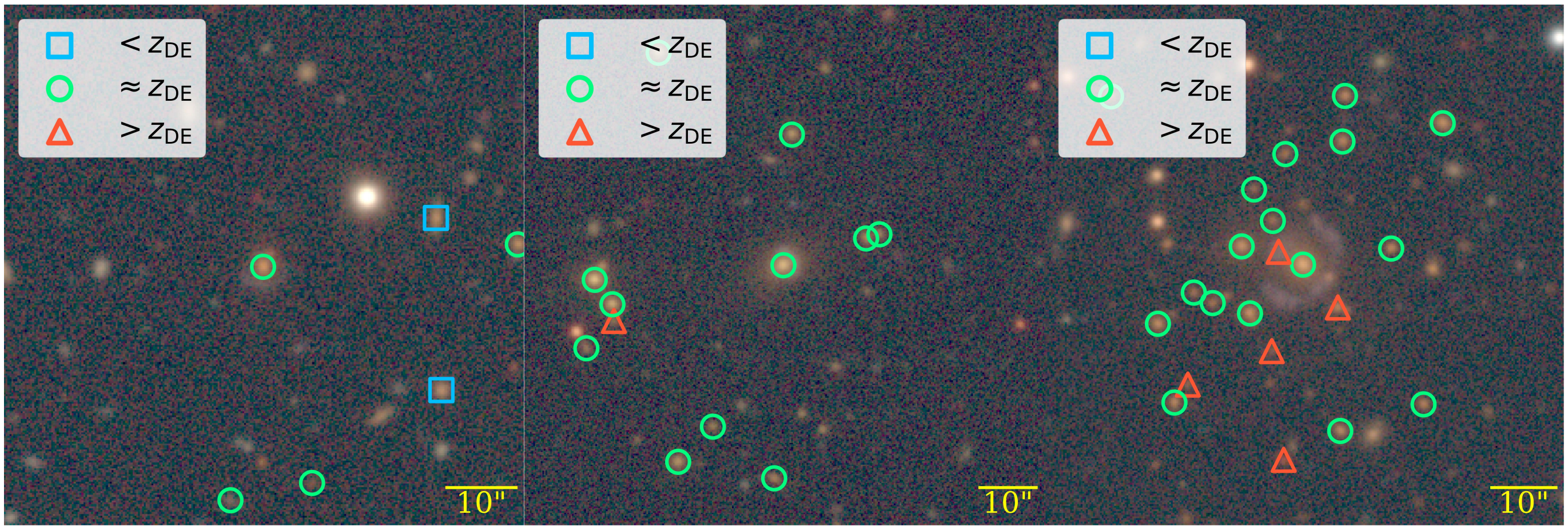

**Figure 4.** Example $z_{\mathrm{phot}}$ measurements of the LOS for (from left) galaxy- (AGEL0003), group- (AGEL0919), and cluster-scale environment candidates. Foreground candidates ($<0.8 \times z_{\mathrm{defl}}$) are shown with blue squares, background candidates ($>1.2 \times z_{\mathrm{defl}}$) are shown with red triangles, and deflector environment candidates are shown with green circles. $z_{\mathrm{phot}}$ and $r$-mag LOS distributions for all 89 systems are available at the Harvard Dataverse: doi: 10.7910/DVN/GYDQQP.

Because of the requirements we place on our photometry, many of the objects with DECaLS photometry (even those visible on cutout images) may not be included in our measurements. In addition, two of our systems, AGEL0243 and AGEL1427, do not have any sources along the LOS fitted to within 20% of $z_{\mathrm{defl}}$. Both lenses are at low ($z < 0.4$) redshifts, so our measurements may result from $z_{\mathrm{phot}}$ bias at low redshifts.

### 3.2.1. $\Sigma_5(z)$ Measurements

Number density measurements are known to be correlated with halo mass (G. O. Abell 1958). Several methods have been implemented to measure environment density, including $r_{200}$-based richness measurements (e.g., S. M. Hansen et al. 2005; M. P. Wiesner et al. 2012), direct $z_{\mathrm{phot}}$ PDF density weighting (S. Taamoli et al. 2024), and the $n$th-nearest neighbor method (M. C. Cooper et al. 2005). M. C. Cooper et al. (2005) find that the $n$th-nearest method most accurately reproduces the environmental density.

To determine the deflector environment density, we apply a redshift-weighted $n$th-nearest neighbor measurement:

$$\Sigma_n(z) = \frac{n}{\pi D_{n,\,\mathrm{Mpc}}^2}(1 + z_{\,\mathrm{defl}}) \tag{3}$$

where $D_{n,\,\mathrm{Mpc}}$ is the distance (in Mpc) between the position of the central deflector and the $n$th deflector environment candidate. $\Sigma_n(z)$ is in units of per $\mathrm{Mpc}^2$. Our $\Sigma_n(z)$ measurements assume that an environment candidate is at $z_{\mathrm{defl}}$, even if it is not fitted precisely at $z_{\mathrm{defl}}$. If the central deflector's $z_{\mathrm{phot}}$ does not fit it to the deflector environment, we use the $z_{\mathrm{spec}}$ to include it as the closest deflector environment candidate. For measurement of the deflector environment density, we choose $n = 5$ to correspond to the fifth innermost member of the deflector environment. We treat the central deflector as one of the five nearest neighbors.

Figure 5 shows a toy model for our $\Sigma_5(z)$ measurements. Using the distances from the center of the field to each fitted environment member (including the main deflector), we measure the distance to the fifth-closest member (labeled "5"), $D_5$. Applying Equation (3), we estimate the environment density.

Because (1) survey incompleteness increases by redshift and (2) the AGEL sample has a wide $z_{\mathrm{defl}}$ distribution, we weigh our $\Sigma_5(z)$ measurements by $1 + z_{\mathrm{defl}}$. Our weighting estimates local environment density even at higher redshifts. Although we do not account for the impact of galaxy evolution on the luminosity function, most early-type galaxies are well assembled by $z = 0.6$, above the median of our sample (D. A. Wake et al. 2006).

For systems with $N_{\mathrm{env}} < 5$ deflector environment candidates, we take the measure $D_{5,\,\mathrm{Mpc}}$ as 250 kpc and $n$ as $N_{\mathrm{env}}$. Our measurement is an upper limit we apply based on the observed number density of the 250 kpc aperture. Our estimates are typically lower than systems with measured $\Sigma_5(z)$.

### 3.2.2. Deflector Environment Classifications

To classify by deflector environment, we apply $\Sigma_5(z)$ cuts based on number density expectations for groups and clusters. We define a galaxy-scale halo as any system where we require upper limit measurements of $\Sigma_5(z)$. In most cases, galaxy-scale deflector environments have multiple halo candidates, but we apply our cut to account for the variability of our $z_{\mathrm{phot}}$ measurements.

We define groups and clusters as systems with $N_{\mathrm{env}}(<250\,\mathrm{kpc}) \geqslant 5$ objects. To separate groups from clusters, we apply a cut equivalent to the $\Sigma_5(z)$ of a deflector at $z_{\mathrm{defl}} = 0.5$ with 15 environment candidates within 250 kpc. We choose $z_{\mathrm{defl}} = 0.5$ as consistent with the median $z_{\mathrm{defl}}$ of our lens sample (see Figure 1). Our choice of 15 deflector environment candidates, on the order of $10^2\,\mathrm{Mpc}^{-2}$, is lower (0.2–0.4 dex) than spectroscopic observations of the cluster of inner ($\lesssim 0.2\,R_{200}$) halos, as shown in Figure 4 of R. F. J. van der

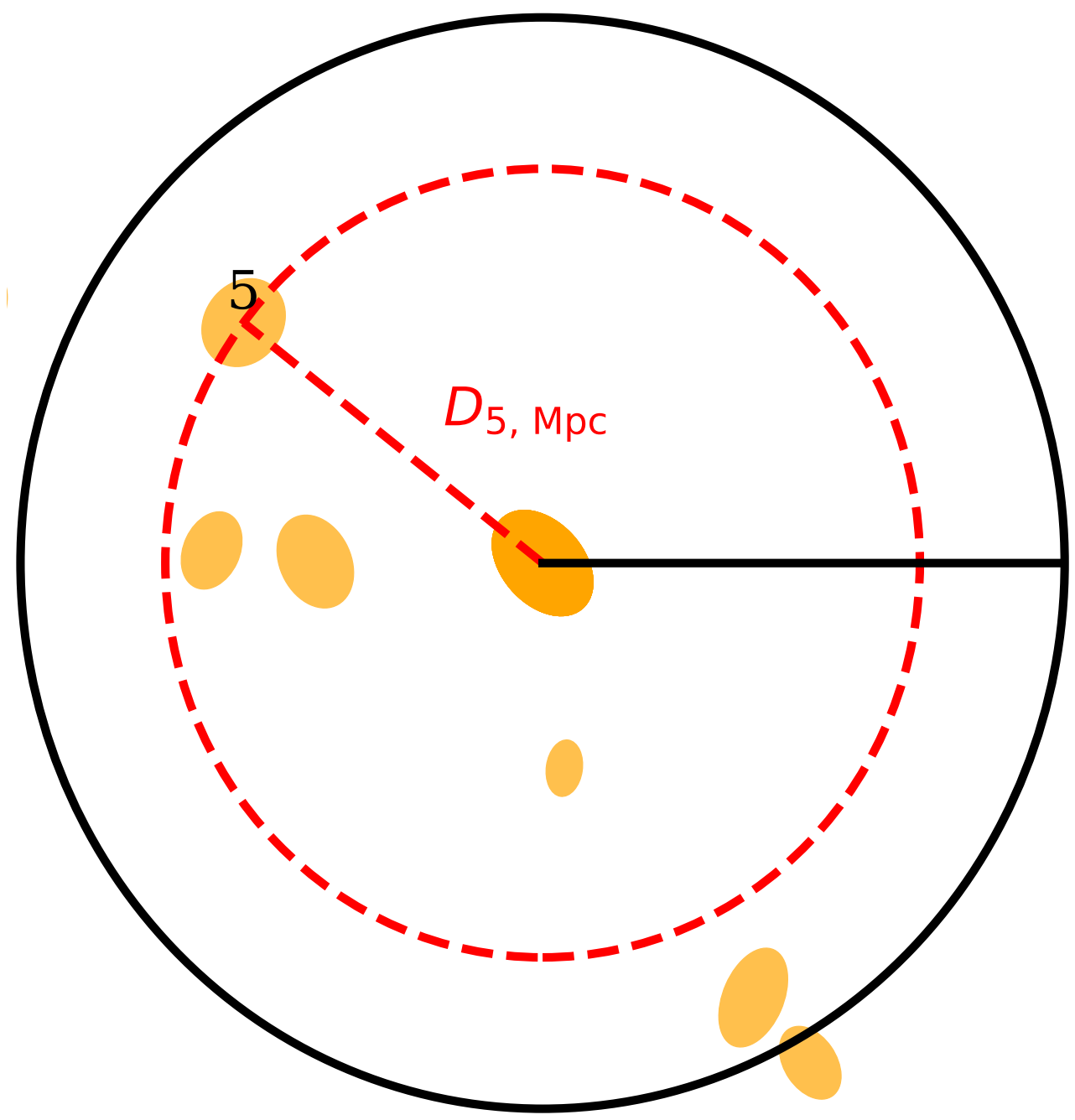

**Figure 5.** Toy model of $\Sigma_5(z)$. The orange ellipses are galaxies fit to the deflector environment at $z_{\mathrm{defl}}$. Including the central deflector, the distance to the fifth-closest fitted deflector environment member (labeled "5") to the center of the field is $D_{5,\ \mathrm{Mpc}}$ (dashed red line). $\Sigma_5(z)$ is the surface density enclosed by the circle with radius $D_{5,\ \mathrm{Mpc}}$ (red, dashed circle).

Burg et al. (2015). However, our underestimate accounts for lower optical depth of our photometry ($r$-band depth of 23.9 in our paper versus 27.7 in R. F. J. van der Burg et al. 2015), our higher-redshift distribution ($z \sim 0.2$–1 versus $z \sim 0.15$), and $z_{\mathrm{phot}}$ unreliability.

#### 3.2.3. LOS Density Comparisons with Control Fields

For each system in our sample, we identify twin galaxies that share characteristics with our central deflectors. Using the twins, we determine whether the AGEL field number densities vary significantly from control fields. To identify twins, we obtain DECaLS DR10 data for a random 16 deg$^2$ field. In that catalog, we isolate galaxies with $\mathrm{SNR}_r \geqslant 5$ and apply morphology cuts, as described in Section 2.2. To match deflectors with twin galaxies, we search for galaxies with the following properties:

1. $r_{\mathrm{defl}} - \delta r \leqslant r_{\mathrm{twin}} \leqslant r_{\mathrm{defl}} + \delta r$, where $\delta r = 0.1$,
2. $(g-r)_{\mathrm{defl}} - \delta(g-r) \leqslant (g-r)_{\mathrm{twin}} \leqslant (g-r)_{\mathrm{defl}} + \delta(g-r)$, where $\delta(g-r) = 0.1$,
3. $z_{\mathrm{defl}} - \delta z \leqslant z_{\mathrm{phot,\ DECaLS}} \leqslant z_{\mathrm{defl}} + \delta z$, where $\delta z = 0.05(1 + z_{\mathrm{defl}})$, and
4. Sérsic (SER) morphology, as classified by Tractor.

With our matching criteria, we find twins for 64 of our 89 systems. We do not observe any systematic differences between the AGEL systems with twins versus those without. The methods we apply to identify control galaxies are similar to C. Faure et al. (2011), who use $z_{\mathrm{phot}}$ and magnitude cutoffs. Because we cannot quantify the reliability of individual DECaLS $z_{\mathrm{phot}}$ measurements, we include $g - r$ cutoffs to provide an additional constraint on redshift. Our requirement of SER morphology ensures consistency with our central deflectors, ~70% of which Tractor classifies as SER.

Due to computational expenses of fitting $z_{\mathrm{phot}}$, we use the photometric redshifts reported by DECaLS ($z_{\mathrm{phot,\ DECaLS}}$), rather than refitting using EAZY, to extract the field size. We use $z_{\mathrm{phot,\ DECaLS}}$ to create an aperture of the radius 250 kpc around each twin galaxy. While we cannot verify the accuracy of our individual $z_{\mathrm{phot,\ DECaLS}}$ measurements, we refer the reader to R. Zhou et al. (2021) for further discussion. Applying the same selection criteria as in Section 2.2, we measure $N_{\mathrm{gal}}$, the total number of galaxies within the aperture, for both the nonlens fields and the AGEL fields. Because we lack twin galaxy $z_{\mathrm{spec}}$ measurements to anchor our control fields to, we cannot reapply the $\Sigma_5(z)$ method as we do for the AGEL fields. Similarly, because our $z_{\mathrm{phot}}$ measurements are biased against higher redshifts, our $N_{\mathrm{gal}}$ measurements are not isolated to deflector environment or LOS objects.

### 3.3. M($<\theta_E$) and Deflector Scale Classifications

For a spherical mass distribution, the Einstein radius, $\theta_{\mathrm{E}}$, is the radius of the lensed image formed by the source (P. Saha et al. 2024). Assuming the source lies directly on the LOS and the deflector is spherically symmetric, we express $\theta_{\mathrm{E}}$ as

$$\theta_{\mathrm{E}}^2 = \frac{4GM}{c^2}\frac{D_{\mathrm{ds}}}{D_{\mathrm{d}}D_{\mathrm{s}}}, \tag{4}$$

where $D_{\mathrm{d}}$, $D_{\mathrm{s}}$, and $D_{\mathrm{ds}}$ are the angular diameter distances to the deflector, to the source, and between the deflector and source, respectively, while $M$ is the mass enclosed by $\theta_{\mathrm{E}}$, called the Einstein mass. For the remainder of this work, we express any reported value of the Einstein mass, $M(<\theta_{\mathrm{E}})$, in $\log_{10} M_{\odot}$.

#### 3.3.1. Measuring $\theta_E$

We use $M(<\theta_{\mathrm{E}})$ measurements to determine the deflector scale. Progress to streamline batch lens modeling is ongoing (e.g., A. J. Shajib et al. 2025), but precise mass modeling of ~$10^2$ deflectors is too computationally expensive for our purposes. Our sample also includes structures with more complex mass distributions than galaxy-scale lenses. Our limitations necessitate simpler methods of measuring Einstein mass.

We use the Markov Chain Monte Carlo (MCMC) algorithm emcee (D. Foreman-Mackey et al. 2013) to estimate $\theta_{\mathrm{E}}$ based on the positions of source images in DECaLS image data. Our MCMC method follows from previous approaches of estimating $\theta_{\mathrm{E}}$ (e.g., J. D. Remolina González et al. 2020). Compared to manual estimates (e.g., using DS9; W. A. Joye & E. Mandel 2003), our method minimizes the impact of human error on our $\theta_{\mathrm{E}}$ estimates. To apply our fit, we visually identify source images and LOS objects in color (DECaLS $g$, $r$, $z$) cutouts. In order to probe only source light, we mask out light from all LOS objects in the cutout. We apply a uniform prior over a range of possible $\theta_{\mathrm{E}}$ that we visually estimate. Our fitting method maximizes the source light contribution per pixel along a ring (the ring model) of the radius $\theta_{\mathrm{E}}$, or in the 9 pixels$^2$ surrounding the end points of a line (the line models) with the half-length $\theta_{\mathrm{E}}$. We apply three $\theta_{\mathrm{E}}$ models based on the source image configuration (as shown in the panels of Figure 6):

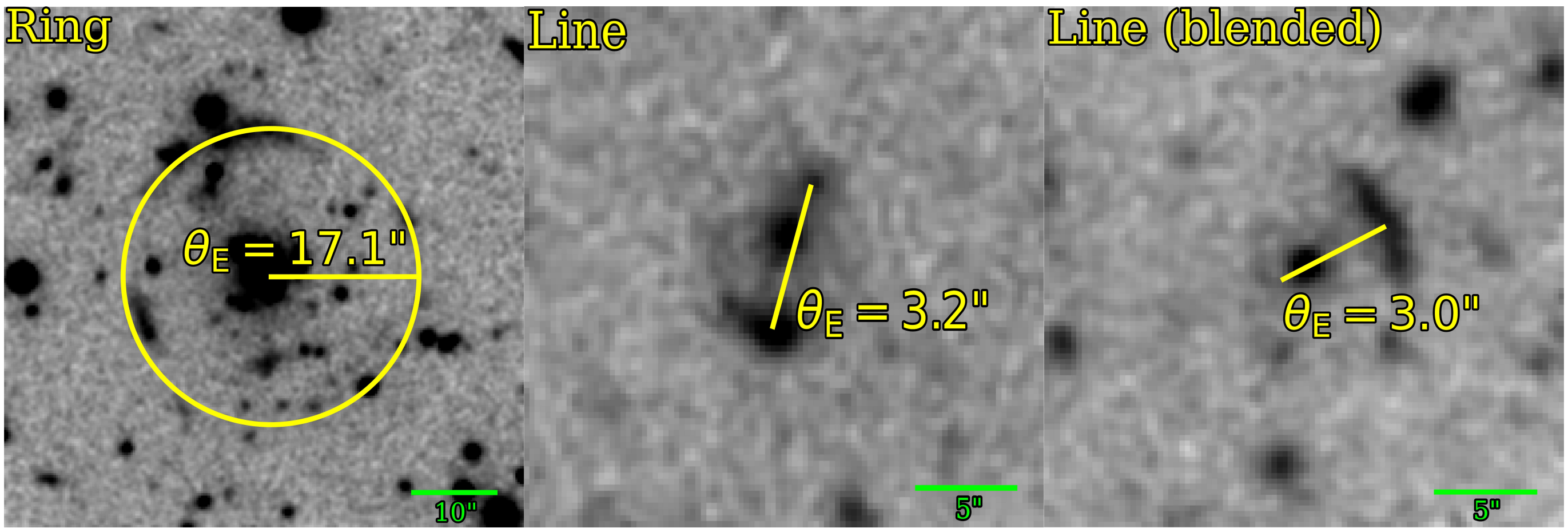

**Figure 6.** $\theta_E$ measurement using a ring model for AGEL0603 ($\theta_E = 17\rlap{.}''1$), line model for AGEL2335 ($\theta_E = 3\rlap{.}''2$), and blended line model for AGEL0911 ($\theta_E = 3\rlap{.}''0$). See Section 3.3.1 for the methodology behind each of these measurements.

1. *Ring model.* If the source images form a ring around the deflector, we fit a ring with the highest source light contribution per pixel. The left panel of Figure 6 shows a ring model fit for AGEL0603. Here, $\theta_E$ is the fitted radius of the ring. We apply this method to 44 systems.
2. *Line model.* If the source image configuration is an image–counterimage pair, we fit a line between the image and counterimage. Our MCMC algorithm maximizes the total source light from the 9 pixel$^2$ region around each end of the line. The center panel of Figure 6 shows an example of a line model fit between an image and counter image in AGEL2335. Here, $\theta_E$ corresponds to half the length of the fitted line. We apply this method to 25 systems.
3. *Line model (blended).* In DECaLS imaging of 16 systems with image–counterimage pairs, the counterimage is blended with deflector light. Here, we estimate the position of the counterimage as the far edge of the deflector light, along the line of axial symmetry with the image. We mask out deflector light and apply a 9 pixel$^2$ region centered at the point we estimate the counterimage to be. We measure $\theta_E$ as half the length of the line fitted between the image and counterimage. The right panel in Figure 6 shows a blended line model fit for AGEL0911. $\theta_E$ is half the length of the fitted line. This method is an upper limit on the estimated $\theta_E$, as the position of the counterimage may be closer to the deflector.

### 3.3.2. Measuring $\theta_E$ Uncertainties

We calculate uncertainties as half the width of the arcs in DECaLS imaging, as measured in DS9 (W. A. Joye & E. Mandel 2003). We find an average arc width of $1\rlap{.}''36$, approximately the size of the $g$-band PSF in DECaLS.

Although uncertainties associated with arc width account for the PSF and pixel scales, there are additional uncertainties based on our assumptions. When the source is directly along the LOS and the deflector has a spherically symmetric mass distribution, the image separation is twice $\theta_E$. In reality, few of the lenses in our sample likely fit these requirements.

To include systematic uncertainties, we measure the scatter between our $\theta_E$ and HST mass models for a subsample of 15 galaxy-scale lenses in N. Sahu et al. (2024) and L. Iwamoto et al. (2026, in preparation; obtained via private correspondence). The lenses are mass modeled using Lenstronomy (S. Birrer & A. Amara 2018) and GLEE (S. H. Suyu & A. Halkola 2010). Figure 7 shows our measurements versus the HST mass-modeled measurements. We measure mean $\frac{\Delta\theta_E}{\theta_E}$ of 0.11, 0.15, and 0.15 for the ring, line, and blended line models, respectively. We attribute the similar $\frac{\Delta\theta_E}{\theta_E}$ between the line and blended line models to a small sample size. We apply the $\frac{\Delta\theta_E}{\theta_E}$ to each of our $\theta_E$ measurements, with the equation

$$\sigma_{\rm sys} = \frac{\Delta\theta_E}{\theta'_E} \times \theta_E, \tag{5}$$

where $\theta'_E$ is the mass-modeled measurement, and $\theta_E$ is our measurement. We choose $\Delta\theta_E$ based on the model we use to measure $\theta_E$.

### 3.3.3. Robustness of Our $\theta_E$ Measurements

For our smaller $\theta_E$ sample ($\theta_E \lesssim 4''$), Figure 7 shows the discrepancies between our measurements and HST mass-modeled measurements for our subsample of 15 deflectors. The low $\frac{\Delta\theta_E}{\theta_E}$ for each model (see Section 3.3.2) suggests that our models accurately reproduce $\theta_E$ at the sample scale.

We observe mean biases of 0.08, 0.08, and −0.006 (weighted by $\theta_E$) for the line, blended line, and ring models, respectively. The positive biases for our line models are expected, as we only measure the separation between the image and counterimage; in many galaxy-scale deflectors, the critical curve is approximately an ellipse, which would have a smaller $\theta_E$ value.

At larger deflector scales, we compare our $\theta_E$ measurement to W. Sheu et al. (2024) for a source at $z_{\rm src} = 1.432$ in AGEL0603. The source image configuration is a quad, so we measure the image separation on each axis and take the average of the two to measure $\theta_E$. Our two $\theta_E$ measurements were $16\rlap{.}''52$ on the long axis and $10\rlap{.}''67$ on the short, yielding $\theta_E = 13.6 \pm 2\rlap{.}''9$. W. Sheu et al. (2024) measures $13.03 \pm 0\rlap{.}''02$, consistent with our measurements. We direct the reader to M. P. Wiesner et al. (2012) and J. D. Remolina González et al.

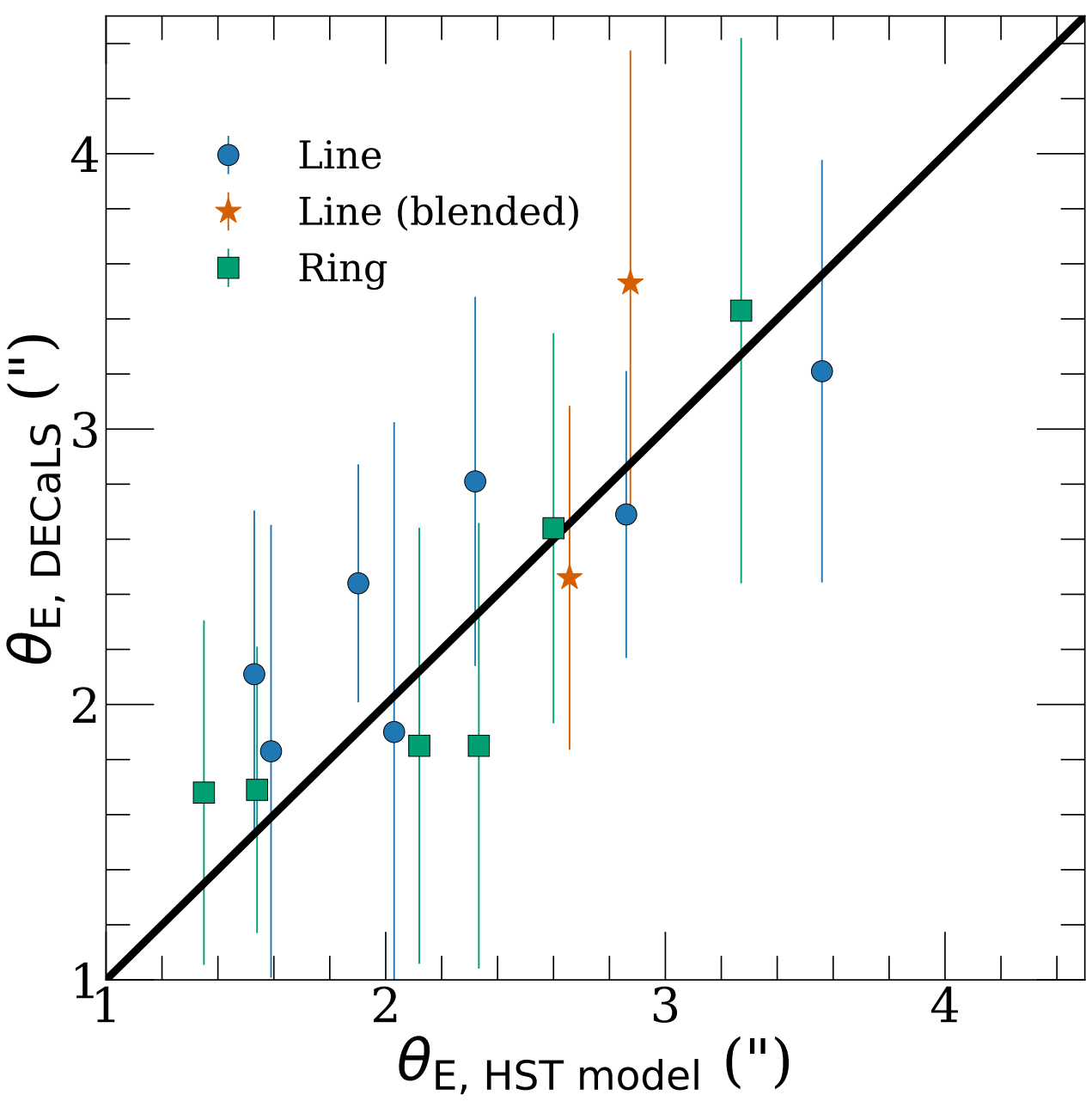

**Figure 7.** Our $\theta_{\mathrm{E}}$ measurements compared to 4 HST photometry mass-modeled lenses in N. Sahu et al. (2024) and 11 in L. Iwamoto et al. (2026, in preparation; obtained via private correspondence). We find $\frac{\Delta\theta_{\mathrm{E}}}{\theta_{\mathrm{E}}} = 0.15, 0.11, 0.15$, for the line (blue circles), circle (green squares), and blended line (orange stars) models, respectively. The error bars correspond to half the primary image arc width measured from DECaLS imaging. We incorporate this scatter into uncertainty for the rest of our $\theta_{\mathrm{E}}$ measurements.

(2020) for data on the robustness of unmodeled group- and cluster-scale $\theta_{\mathrm{E}}$ measurements.

#### 3.3.4. M( $< \theta_E$) Measurements and Deflector Scale Classifications

For each system with measured $\theta_{\mathrm{E}}$, we find $M( < \theta_{\mathrm{E}})$ by solving for $M$, the Einstein mass, in Equation (4). We estimate the $M( < \theta_{\mathrm{E}})$ for 85 lenses. Our $M( < \theta_{\mathrm{E}})$ measurements range from $\log \frac{M(< \theta_{\mathrm{E}})}{M_{\odot}} \sim 11.8 - 14.0$. We measure a median Einstein mass of $\log_{10}\left(\frac{M(< \theta_{\mathrm{E}})}{M_{\odot}}\right)_{\mathrm{avg}} = 12.30 \pm 0.27$. The diverse range of $M( < \theta_{\mathrm{E}})$ in our sample provides a simultaneous look at lenses across the Einstein mass spectrum.

For six of the galaxy-scale lenses in N. Sahu et al. (2024), they report $\overline{M( < \theta_{\mathrm{E}})} = 12.1 \pm 0.3$. For the same six lenses, we estimate a mean $\overline{M( < \theta_{\mathrm{E}})} = 12.2 \pm 0.3$, consistent within 0.1 dex of the N. Sahu et al. (2024) measurements. At higher-mass scales, we compare the enclosed mass measured from our $\theta_{\mathrm{E}}$ measurement of the $z_{\mathrm{src}} = 1.432$ source from W. Sheu et al. (2024). We find $M( < \theta_{\mathrm{E}}) = 13.70 \pm 0.19$, similar to the measurements reported in W. Sheu et al. (2024). J. D. Remolina González et al. (2020) finds that $\theta_{\mathrm{E}}$ accurately reproduces $M( < \theta_{\mathrm{E}})$ at larger Einstein mass scales.

To classify lenses according to their deflector scale, we apply deflector mass cuts. Assuming a singular isothermal ellipsoid (SIS) deflector, $\theta_{\mathrm{E}}$ is given by

$$\theta_{\mathrm{E}} = 4\pi \frac{\sigma^2}{c^2} \frac{D_{\mathrm{ds}}}{D_{\mathrm{s}}}, \tag{6}$$

where $\sigma$ is the velocity dispersion. If we define $\Sigma_{\mathrm{cr}} = \frac{c^2 D_{\mathrm{s}}}{4\pi G D_{\mathrm{ds}} D_{\mathrm{d}}}$, then the 2D enclosed mass is

$$M(<\theta_{\mathrm{E}}) = \pi \Sigma_{\mathrm{cr}} (\theta_{\mathrm{E}} D_{\mathrm{d}})^2. \tag{7}$$

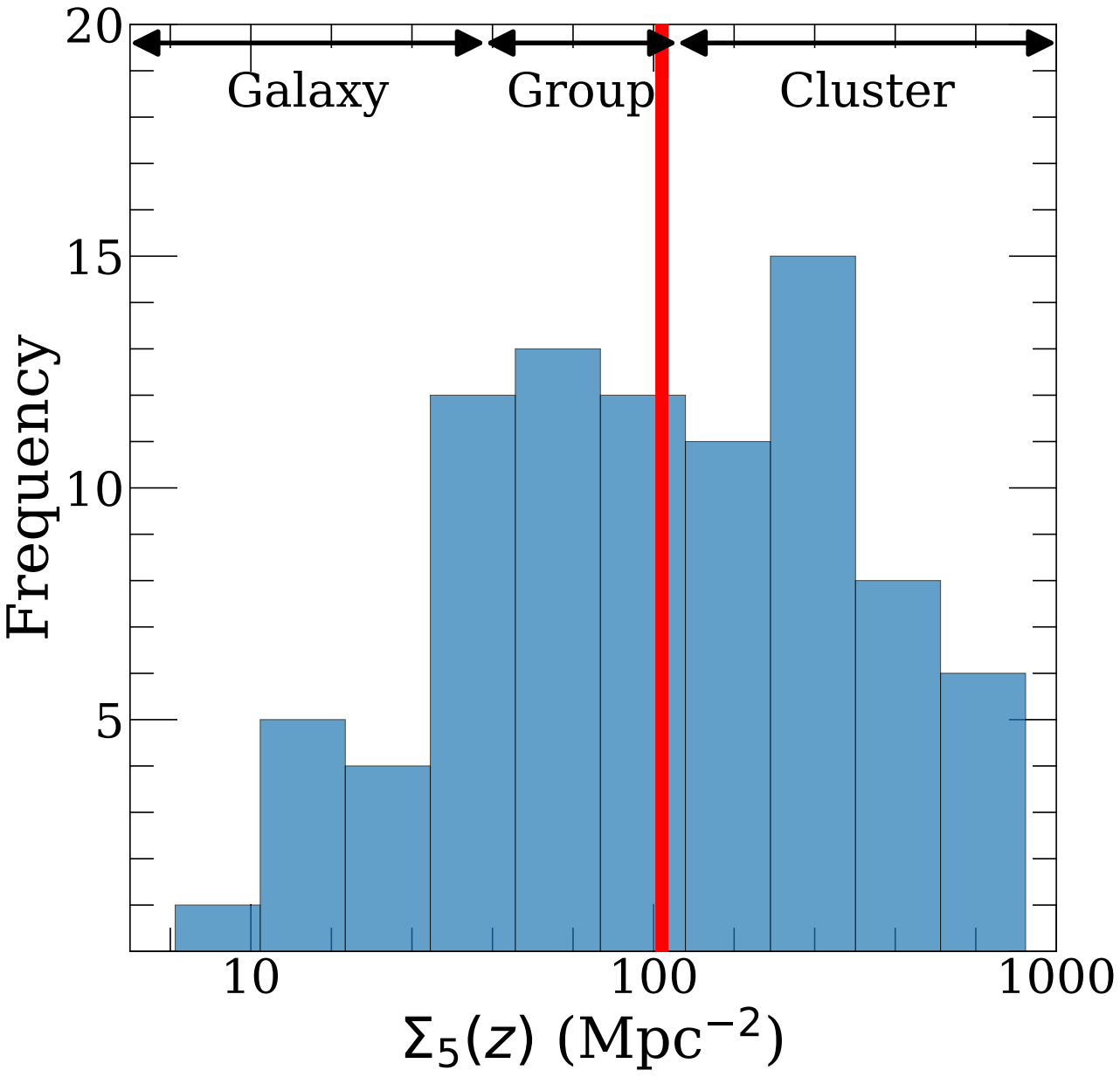

**Figure 8.** $\Sigma_5(z)$ distribution for 87 lenses. The red line denotes the mean $\Sigma_5(z)$ for our sample (104.9 $\mathrm{Mpc}^{-2}$; which corresponds to a large group). Our distribution shows that AGEL survey deflectors are dominated by group and cluster halos. We include galaxy-scale deflector environments, which have upper limit measurements on $\Sigma_5(z)$ (see Section 3.2.1).

The proportion of mass enclosed by an SIS profile is sensitive to velocity dispersion (i.e., halo mass) and redshifts of the deflector and source. For high-mass halos at our typical $z_{\mathrm{defl}}$ and $z_{\mathrm{src}}$, $M( < \theta_{\mathrm{E}}) \sim 0.1 M_{200}$, where $M_{200}$ is the halo mass. Therefore, to apply our cuts, we assume that $M( < \theta_{\mathrm{E}})$ is 10% of the total halo mass.

We define (log) $M_{200} < 13$ as galaxies, 13–14 as groups, and $>14$ as clusters. Therefore, using the $M( < \theta_{\mathrm{E}})$ measurements, we classify deflectors with $M( < \theta_{\mathrm{E}}) < 12$ as galaxy-scale deflectors, 12–13 as group-scale deflectors, and $>13$ as cluster-scale deflectors. Our values are consistent with the expected $M( < \theta_{\mathrm{E}})$ at the mass scales of galaxy, group, and cluster halos.

When we cannot make reliable measurements of $\theta_{\mathrm{E}}$ (for six systems), we make visual classifications of the deflectors using DECaLS photometry. If DECaLS photometry only shows one deflector enclosed by $\theta_{\mathrm{E}}$, then we classify the deflector as galaxy-scale. If DECaLS photometry shows multiple deflectors, we classify the deflector as group-scale.

## 4. Results

### 4.1. $\Sigma_5$(z) Distribution and Deflector Environment Classifications

Figure 8 shows the $\Sigma_5(z)$ distribution for 87 lenses from our original sample. We exclude two systems with zero fitted deflector environment objects. The vertical red line is the mean $\Sigma_5(z)$ of our sample (104.9 $\mathrm{Mpc}^{-2}$, corresponding to a dense group). Interestingly, our measurements show that most of the AGEL survey lenses are in group and cluster environments. The lack of galaxy-scale environments is in disagreement with the halo mass predictions of QSOs and SNe in K. T. Abe et al. (2025), who predict that lens detections should peak at the galaxy-scale—$\log_{10} M_{\mathrm{deflector\ halo}} \approx 12$. Given the AGEL sample's complex selection function, we cannot make any conclusions as to the *true* observed halo mass distribution of

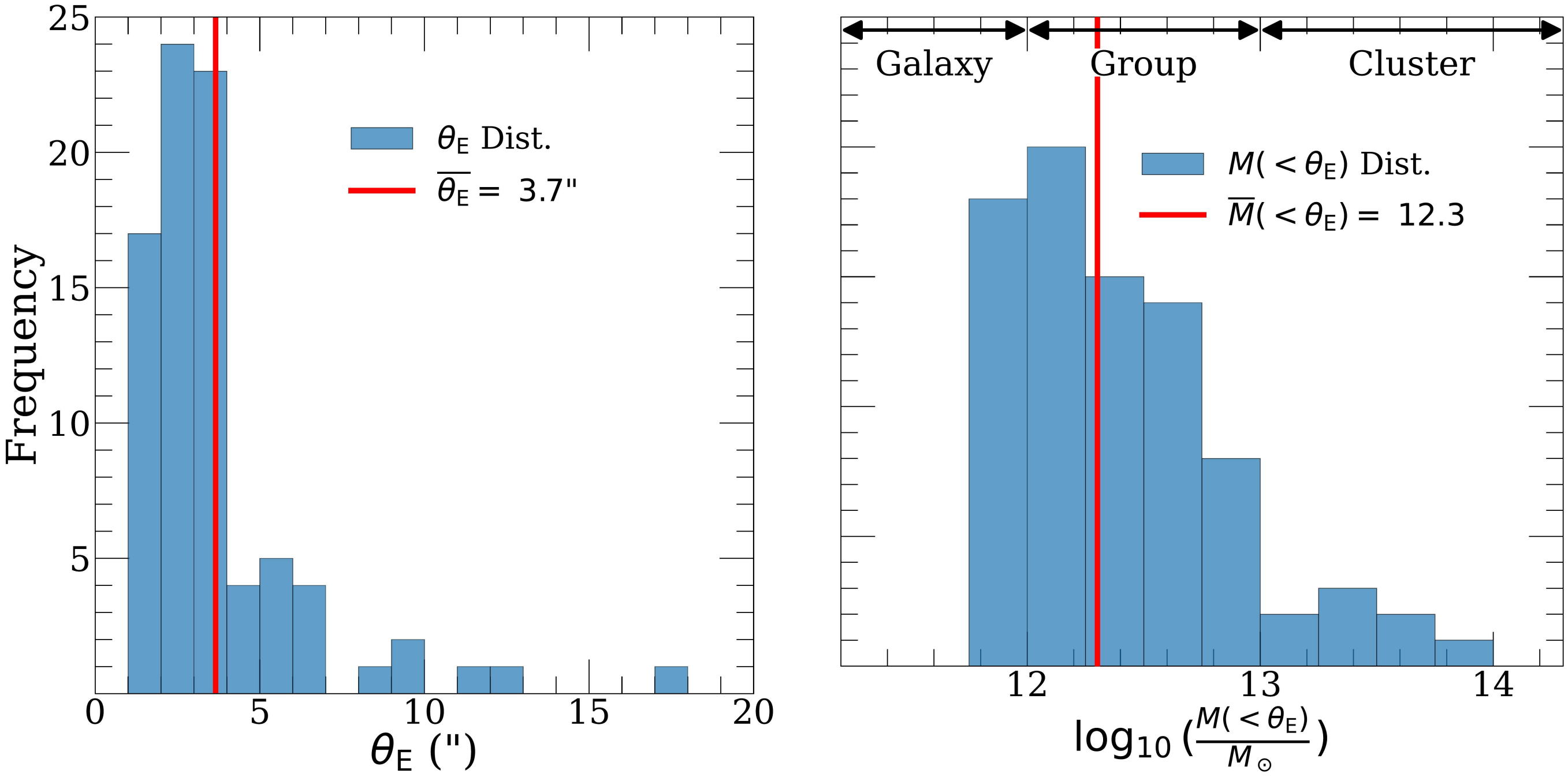

**Figure 9.** Left: $\theta_{\mathrm{E}}$ distribution for 85 lenses in the AGEL sample. We measure a mean $\theta_{\mathrm{E}}$ (red solid line) of 3$''$7. Right: $M(<\theta_{\mathrm{E}})$ distribution for the same 85 lenses. We measure a median $M(<\theta_{\mathrm{E}})$ of 12.3 (in $\log_{10} M/M_{\odot}$), consistent with a group-scale Einstein mass.

**Table 1**
Left *Five Columns*: Deflector Environment Classification Distributions for the 89 Systems for Which We Made $z_{\mathrm{phot}}$ Measurements

| Env. Scale | $\Sigma_5(z)$ Cutoff | Frequency | $\bar{z}_{\mathrm{defl}}$ | $\overline{M}(<\theta_{\mathrm{E}})$ | Deflector Scale | $\log_{10} M(<\theta_{\mathrm{E}})$ Cutoff | Frequency | $\bar{z}_{\mathrm{defl}}$ | $\overline{M}(<\theta_{\mathrm{E}})$ |
|---|---|---|---|---|---|---|---|---|---|
| Galaxy | No measurement | 21 | 0.53 | 12.17 ± 0.28 | Galaxy | <13 | 81 | 0.54 | 12.24 ± 0.28 |
| Group | Measurement | 26 | 0.53 | 12.18 ± 0.29 | Group | 13–14 | 8 | 0.51 | 13.44 ± 0.19 |
| Cluster | $\gtrsim$115 | 42 | 0.55 | 12.52 ± 0.26 | Cluster | $\geqslant$14 | 0 | N/A | N/A |

**Note.** We provide median $M(<\theta_{\mathrm{E}})$ (in $\log_{10} M/M_{\odot}$) measurements for each environment scale. Right five columns: deflector scale classification distributions for all 89 systems. Our $M(<\theta_{\mathrm{E}})$ averages by deflector scale are inconsistent with associated averages by environment scale, suggesting a disconnect between lens scale and lens environment.

deflectors in the Universe. Additional data, particularly spectroscopic follow-ups of AGEL LOS and development of less-biased spectroscopic samples, are necessary to confirm our finding.

Using our $\Sigma_5(z)$ measurements to classify deflector environment, we find 21 systems with galaxy-, 26 with group-, and 42 with cluster-scale deflector environments. The proportion of cluster deflector environments is much higher than mentioned in T. Treu et al. (2009). However, the T. Treu et al. (2009) sample includes only galaxy–galaxy lenses. Their cluster count also only includes clusters confirmed in the literature. Our higher proportion of clusters may be due to (1) the AGEL survey's lower limit ($\sim$2$''$) on estimated $\theta_{\mathrm{E}}$ and (2) the more massive halos required to produce highly magnified sources. In addition, the physics of strong lensing favors larger (and denser) halos (K. T. Abe et al. 2025). We discuss that result further in Section 5.2.

The first five columns of Table 1 show the $\Sigma_5(z)$ cutoffs, frequencies, and median $z_{\mathrm{defl}}$ of our three environment scales. We measure a median $z_{\mathrm{defl}}$ of 0.53 for galaxy-, 0.53 for group-, and 0.55 for cluster-scale environments. The lack of redshift-dependence on our deflector environment scales is consistent with the expectation that early-type galaxies and their host halos evolve only passively at low redshifts (e.g., E. O. Ofek et al. 2003; M. Oguri 2006).

### 4.2. $\theta_E$ Distribution and Deflector Scale Classifications

The left panel of Figure 9 shows the $\theta_{\mathrm{E}}$ distribution for 85 systems. The mean $\theta_{\mathrm{E}}$ of our distribution is 3$''$7. Most of our sample has measured $\theta_{\mathrm{E}} < 5''$. Our distribution peaks at approximately 2$''$–3$''$, whereas predictions in K. T. Abe et al. (2025) peak at $\sim$1$''$. C. Jacobs et al. (2019a) apply a $\theta_{\mathrm{E}}$ cutoff of 2$''$ for simulated lenses in their training sets, which places a lower limit on $\theta_{\mathrm{E}}$. The AGEL selection is also limited by the DES and DECaLS PSFs.

The right panel of Figure 9 shows the $M(<\theta_{\mathrm{E}})$ distribution. We measure a median (log) $M(<\theta_{\mathrm{E}})$ of 12.3, consistent with the mass of a large galaxy (e.g., R. Mandelbaum et al. 2006), and assuming an SIS, consistent with a small $M_{200} \sim 13$ group. From our distribution, most of the AGEL survey lenses have group-scale Einstein masses. As shown in the five right columns of Table 1, we classify 22 systems as galaxy-, 59 as group-, and 8 as cluster-scale lenses. Cluster-scale lensing is predicted to be quite rare in future large-scale surveys (K. T. Abe et al. 2025), so the small proportion of cluster-scale lens detections is not surprising. Nonetheless, the small number of cluster-scale lenses suggests that there may also be an upper limit on $\theta_{\rm E}$ (and implicitly Einstein mass) in the AGEL sample. We explore this possibility further in Section 5.

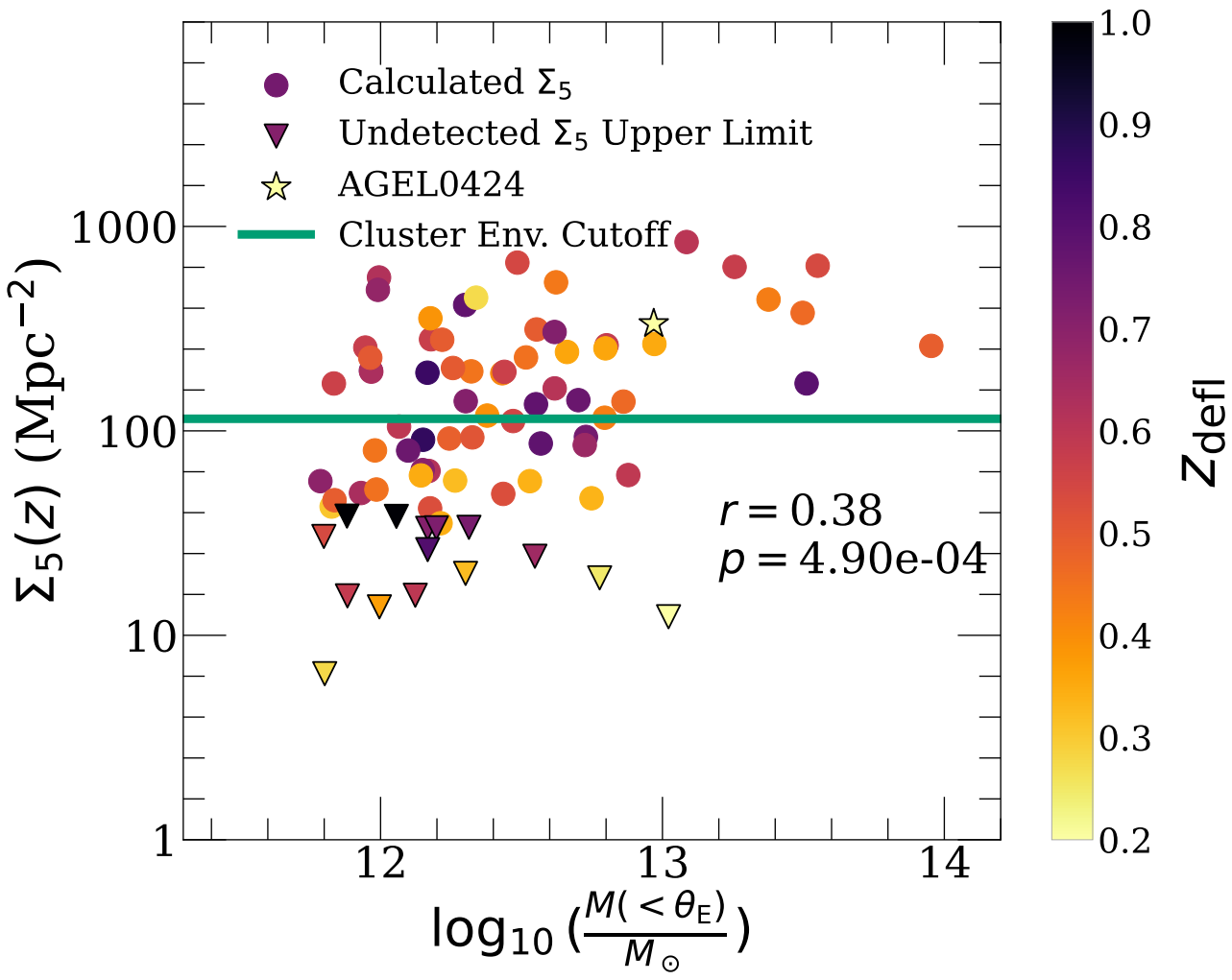

**Figure 10.** $\Sigma_5(z)$ versus measured Einstein mass ($\log_{10}\frac{M(<\theta_{\rm E})}{M_\odot}$). The downward-facing triangles are systems for which we estimate the upper limit $\Sigma_5(z)$, the circles are systems with measured $\Sigma_5(z)$, and the star is the system AGEL0424 (see Figure 11). The color scheme is based on $z_{\rm defl}$. We find a Spearman correlation of 0.38, suggesting only a weak correlation between Einstein mass and halo mass.

## 5. Discussion

### 5.1. Einstein Mass Does Not (Reliably) Trace Halo Mass

Table 1 shows the statistics of deflector scale and environment classifications for our sample. Whereas 47.2% of deflectors are in cluster-scale environments, only 9.0% have cluster-scale Einstein masses. In addition, the median deflector in a cluster environment has Einstein mass corresponding to a group-scale deflector. Our results suggest that the deflector scale is not closely correlated with deflector environment, especially in the densest (cluster) environments.

Figure 10 shows $\Sigma_5(z)$ (our tracer of deflector environment) versus $M(<\theta_{\rm E})$ (our tracer of deflector scale) for 85 lenses. The circles are systems with measured $\Sigma_5(z)$, the triangles are systems with upper limit estimates of $\Sigma_5(z)$, and the star is AGEL0424 (see Figure 11). The color scheme is based on $z_{\rm defl}$. For all systems (including those with $\Sigma_5(z)$ upper limits), we measure a Spearman correlation coefficient of $r = 0.38$. For only systems with measured $\Sigma_5(z)$, we find $r = 0.36$. The weak correlation between the deflector scale and environment suggests that the deflector scale does not reliably trace deflector environment.

A few possible explanations exist for the lack of a strong observed correlation. First, some deflectors may be satellites in larger halos. In several systems—particularly some of our galaxy-scale deflectors in cluster environments—we visually identify nearby, brighter galaxies with similar colors to our deflector. Dense cluster environments boost the lensing probability for satellites relative to field galaxies with the same properties, so our result is not unexpected. Nonetheless, spectroscopic or X-ray observations are needed to confirm whether some AGEL deflectors are satellites.

In addition, AGEL detections may be biased toward systems with high source magnifications, $\mu$. K. T. Abe et al. (2025) predicts that, for $\mu \gtrsim 10$, the host halo mass distribution is roughly flat for $M_{200} \sim$ 12–14. Coupled with AGEL's bias toward higher-$\theta_{\rm E}$ fields, we would expect more systems in group and cluster halos, consistent with Figure 8. Similarly, because the C. Jacobs et al. (2019a) CNN is trained on galaxy-scale lenses, we expect a bias toward higher-magnification, galaxy- and group-scale lenses.

From Equation (4), there is a redshift dependence on $\theta_{\rm E}$—and thus $M(<\theta_{\rm E})$. To investigate the impact of cosmology on our lack of correlation, we isolate our sample to deflectors with $0.5 < z_{\rm defl} < 0.6$ and $2 < z_{\rm src} < 3$. Our method isolates our $\Sigma_5(z)$ versus $M(<\theta_{\rm E})$ for systems with roughly (within ∼20%) equivalent $\theta_{\rm E}$ dependence on cosmology. For seven systems, we measure a Spearman coefficient of $r = 0.5$. Although we find a slightly higher correlation when adjusting for cosmology, cosmology does not explain the observed lack of a correlation for our full sample. Per Table 1, we do not observe a redshift dependence in our deflector scale classifications, suggesting that $M(<\theta_{\rm E})$ is the primary driver of $\theta_{\rm E}$ in the AGEL sample.

The lack of correlation may be closely related to AGEL survey selection effects, as the training sets behind the AGEL sample may present an upper limit on $\theta_{\rm E}$. As shown in Figure 1 in C. Jacobs et al. (2019a), the training set images for the AGEL selection are galaxy-scale lenses. Because the CNNs are trained using galaxy-scale lenses, larger-$\theta_{\rm E}$ features, such as a giant arch, may be less effectively detected in the AGEL selection. To achieve more representative samples for a batch analysis of galaxy clustering and cosmology, future lens searches should incorporate group- and cluster-scale lensing in their training sets.

Another selection effect results from observational limitations of the DECaLS fields. Figure 11 shows AGEL0424, a galaxy-scale lens in a cluster-scale environment. The left panel shows a red, green, and blue (RGB; $z$-, $r$-, $g$-band) image from the DECaLS observations while the right panel shows a false-color image from follow-up HST observations (see Section 2). The HST follow-up reveals two source candidates lying 22.″5 and 26.″7 from the central deflector. In the left panel, we circle the two source candidates in corresponding DECaLS data. Here, the sources are visible, but are too unresolved to be detected as lensed sources. We also find source candidates with $\theta_{\rm E} > 30''$ in AGEL0603, and we suspect that many other such sources would be visible in HST follow-up observations of our cluster-scale environment sample.

Assuming a redshift of $z = 1.87$ (the average $z_{\rm src}$ from our sample), the $M(<\theta_{\rm E})$ corresponding to source 2 in AGEL0424 would be $M(<\theta_{\rm E}) \approx 14.3$—a large cluster-scale deflector. Our value is 1.3 dex greater than the $M(<\theta_{\rm E})$ we measure for AGEL0424 ($M(<\theta_{\rm E}) = 12.97$). As AGEL0424 is in a cluster-scale deflector environment, our higher-$\theta_{\rm E}$ measurements support a stronger correlation between the deflector scale and environment. In the absence of follow-up observations, we cannot confirm this selection effect for most of the systems located in dense deflector environments. However, the presence of multiple arcs in HST follow-ups of two clusters hints that many more such arcs may be visible in our cluster environment sample.

### 5.2. AGEL Fields versus Control Fields

Figure 12 shows the $N_{\rm gal}$ distribution of AGEL fields (in blue) and control fields (in orange; normalized to size of AGEL

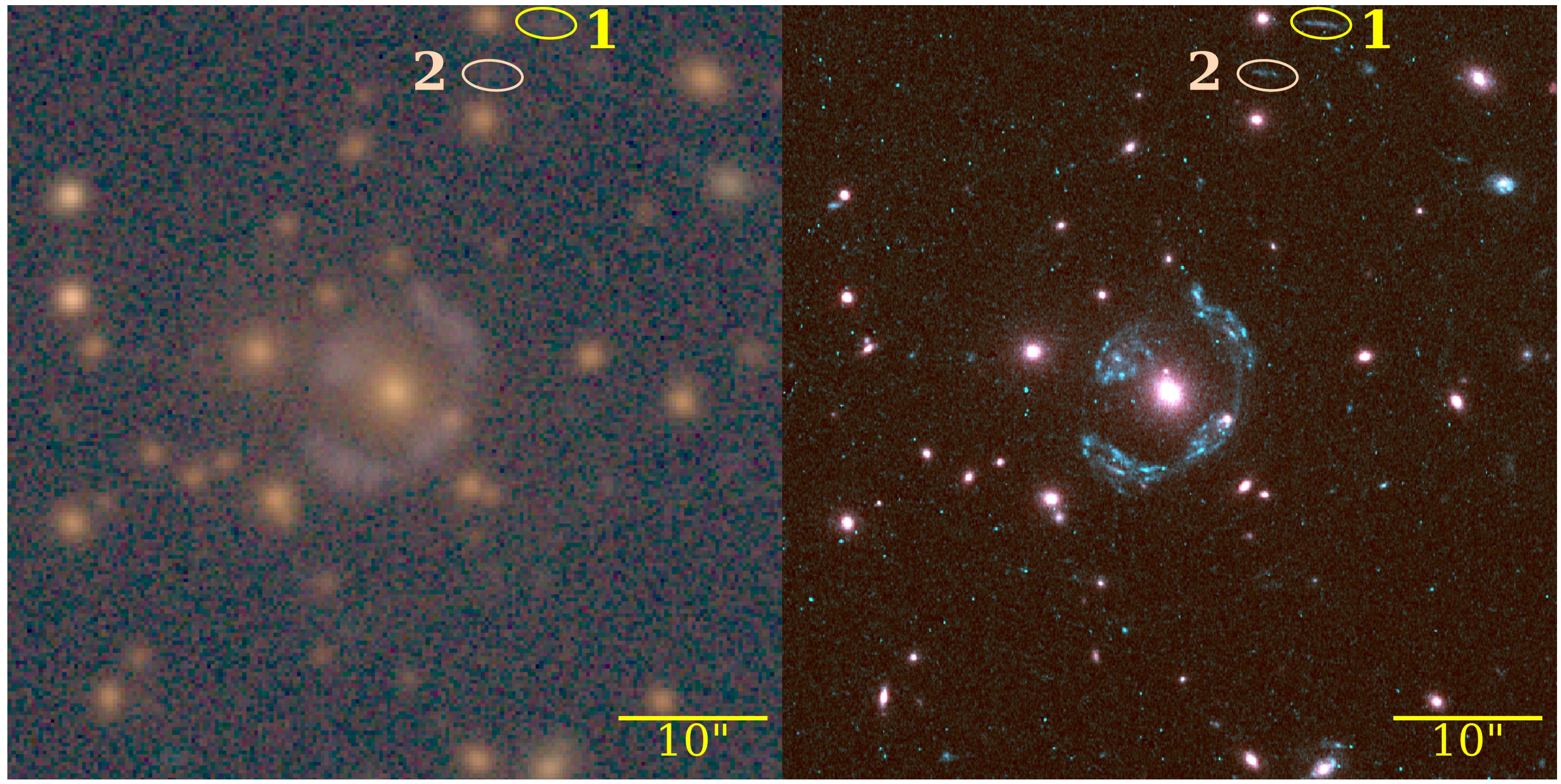

**Figure 11.** Field-of-view images of AGEL0424, which is classified as a galaxy-scale deflector in a cluster-scale deflector environment. Left: DECaLS 52″ × 52″ image of the line of sight; the labels 1 and 2 enclose source candidates detected in HST data but undetected in DECaLS data. Right: deeper HST imaging in WFC3 F200LP and F140W bands reveals fainter sources (labeled 1 and 2) not readily visible in ground-based DECaLS imaging. Absent high-resolution imaging, the additional source candidates on the right figure are not visible in large-scale surveys. We measure the angular distance from the arcs to the central deflector as 22.″5 and 26.″7, respectively. Absent observations of sources with significant image separations, many cluster deflecting halos may have lensing configurations that cannot trace halo mass without consideration of the mass environment of the deflector plane.

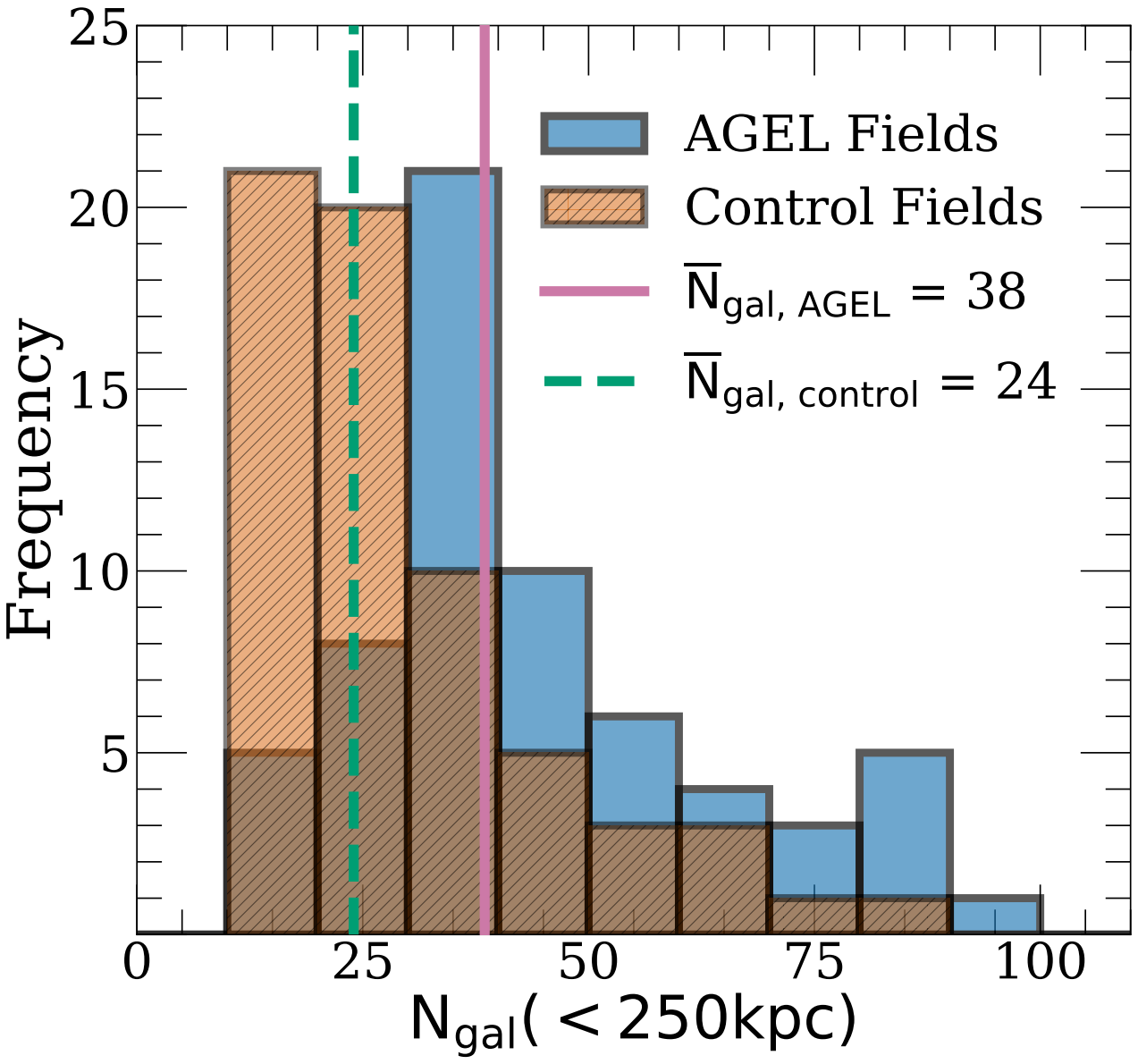

**Figure 12.** Number of objects with measured $z_{phot}$ inside 250 kpc radius (proper coordinates) apertures for AGEL fields (in blue) and control fields we select using similar selection methods as in C. Jacobs et al. (2019a; in orange). The median $N_{gal}(<250\,kpc)$ for AGEL fields (pink solid line) and control fields (green-dotted line) are 38 and 24, respectively, suggesting that AGEL fields are located in systematically denser environments than control fields.

distribution). The solid pink and dashed green lines are the median $N_{gal}(<250\,kpc)$ for the AGEL and control fields, respectively. We measure a median of 38 objects in our AGEL fields versus 24 for our control fields. We apply a Kolmogorov–Smirnov (KS) test to determine whether the AGEL and control fields are distinguishable distributions, and we find a vanishingly small ($\sim 10^{-6}$) $p$. Our results show significant differences between the $N_{gal}$ of AGEL versus control fields.

Our result is inconsistent with several past studies comparing lensing fields with control fields (e.g., T. Treu et al. 2009; C. Faure et al. 2011; K. C. Wong et al. 2018). The inconsistency may be due to differences between the AGEL sample and samples used in past results. For example, the T. Treu et al. (2009) analysis is based on the Sloan Lens Advanced Camera for Surveys (A. S. Bolton et al. 2006), which probes lenses primarily with $\theta_E \lesssim 2''$ (A. S. Bolton et al. 2008). Similarly, the C. Faure et al. (2011) and C. D. Fassnacht et al. (2011) analyses are based on the COSMOS field lenses (C. Faure et al. 2008), which also have $\theta_E \lesssim 2''$. Compared to other surveys, our $\theta_E$ distribution is biased toward much larger deflectors (see Figure 9).

Relative to twin galaxies chosen with the same individual properties, AGEL deflectors are biased toward group and cluster environments. Our result is consistent with our other measurements, which show that the AGEL deflector sample is strongly biased toward group-scale Einstein masses and cluster-scale environments. As the Figure 10 result suggests, many of our deflectors may be satellites in groups and clusters. As a result, we would expect that AGEL fields would be located in denser environments than control fields.

AGEL lenses may also be located in denser LOSs (i.e., outside the deflector environment) than control fields. However, because of computational constraints, we cannot confirm whether some of the observed discrepancy is due to an excess of objects outside the deflector environment. Both M. W. Auger (2008) and C. D. Fassnacht et al. (2011) find insignificant bias in LOS density between lensing and control fields. Because a majority of objects with EAZY$z_{\rm phot}$ are classified as deflector environment candidates, we do not expect to observe significant overdensity outside the deflector environment. Nonetheless, more work is needed to confirm whether AGEL systems are biased toward denser LOSs.

The difference between AGEL and control fields is sensitive to the $r$-band magnitude distribution of our AGEL deflectors. If we match twins only for AGEL deflectors with $r < 19$, we find the median $N_{\rm gal}$ of $\overline{\rm N}_{\rm gal,\,AGEL} = 50$ and $\overline{\rm N}_{\rm gal,\,control} = 42$, slightly decreasing the observed difference. Our KS probability also improves to $p = 0.05$, although some constraining power is lost due to the smaller ($\sim$20) sample. The KS $p$-value still concludes to $\sim 2\sigma$ that the AGEL and control distributions are different. We hypothesize that fainter (i.e., $r > 19$) control galaxies may have less reliable $z_{\rm phot,\ DECaLS}$ measurements, which may lead to imprecise deflector-twin matching. Furthermore, at the brightest $r$-band magnitudes, we are much more likely to choose clusters both for the AGEL and control fields. Nonetheless, neither bias accounts for most of the observed difference between AGEL and control fields.

Theoretical predictions of gravitational lensing phenomena show that the probability of finding strong-lensing events is higher for larger halos (e.g., C. Saez et al. 2016), as well as subhalos in larger halos (M. Meneghetti et al. 2020). As larger halos have larger lensing cross sections than smaller halos, larger halos are associated with higher probability of lensing. Our observation of the high $N_{\rm gal}$ tail is consistent with the higher proportion of cluster-scale environments in the AGEL sample. Observations of cluster substructures also show that cluster-scale halos are more efficient lenses than $\Lambda$CDM predicts by more than an order of magnitude (M. Meneghetti et al. 2020). Although an analysis of the discrepance is beyond the scope of our paper, it is an interesting result that warrants further investigation. Our observations provide a method to analyze this phenomenon with future, less biased samples.

### *5.3. Accounting for Deflector Environment in Lens Modeling*

Our $z_{\rm phot}$ and $\theta_{\rm E}$ measurements provide promising and efficient methods to (1) identify systems of interest and (2) minimize systematics due to environment in lens modeling. New surveys from Euclid and LSST each are expected to discover $\mathcal{O}(10^5)$ galaxy-scale lenses (T. E. Collett 2015). We have shown that our simple $z_{\rm phot}$ and $\theta_{\rm E}$ measurement methods are useful to identify lenses with sparse LOSs or special characteristics (e.g., group-scale lenses or galaxy-scale lenses in clusters). Our method could improve systematics in lens modeling and improve lens selections for a host of science applications.

M. L. Wilson et al. (2017) find that external convergence from dense deflector environments can lead to measurements overestimating $H_0$ by $\sim$10%. Similarly, they find that precise measurements of $H_0$ require accounting for nearby structures along the LOS. Therefore, to make one of the most precise late Universe measurements of $H_0$ yet, the H0LiCOW team in K. C. Wong et al. (2019) uses wide-field spectroscopic data of the deflector environment to capture the LOS mass distribution. Although we lack spectroscopic redshifts of the environment in our sample, our $z_{\rm phot}$ measurements provide a method to roughly account for environment in lens models. Our methods can also be used to identify samples for a simultaneous analysis of halo characteristics across the mass spectrum.

In the future, as new spectroscopic surveys such as the Dark Energy Spectroscopic Instrument (DESI; DESI Collaboration et al. 2025) provide spectroscopic measurements of the LOS for our sample, we could further improve systematics in lens modeling by using spectroscopic, rather than photometric redshifts. We suggest a repetition of our analysis once sufficient spatial spectroscopic data are available.

Although we do not provide constraints on individual masses of LOS members, we can use $r$-band magnitudes as a tracer. To that end, we provide $r$-band magnitude maps for all objects with measured $z_{\rm phot}$ along the LOS in each system. In addition to maps, we provide tables of coordinates, fitted $z_{\rm phot}$, and $r$-mag for all objects fitted in the LOS for each system. All of our data are available as a supplement to our paper, and will be made public via the AGEL (see footnote 12) website.

We investigate whether accounting for environment using our lens modeling prescription improves mass models for AGEL0424, a group-scale deflector located in a dense cluster halo. A further analysis of this lens, which has been dubbed "Gargantua," will appear in L. Iwamoto et al. (2026, in preparation). Using GLEE (S. H. Suyu & A. Halkola 2010) without inclusion of the deflector environment in position models, Gargantua does not converge on a lensing solution. We use our $r$-mag and $z_{\rm phot}$ measurements to apply a host halo profile, and we test whether this improves our position model in GLEE. We find that accounting for environment leads to convergence and substantial qualitative improvement in image positions. Quantitative results of the position model will be published in L. Iwamoto et al. (2026, in preparation). Our measurements demonstrate that our rough method to account for deflector environment may substantially improve lens models.

## 6. Conclusions

In our work, we measure the relation between the deflector scale (Einstein radius) and deflector environment (galaxy surface density) for a sample of 89 spectroscopically confirmed strong lenses from the AGEL survey. The deflectors are at $z \sim 0.2$–1, and the sources are at $z \sim 1$–3 (see Figure 1 for $z_{\rm spec}$ distribution). To quantify the halos of the deflectors, we use EAZY (G. B. Brammer et al. 2008) to measure photometric redshifts ($z_{\rm phot}$) for all objects along the LOS and use the $z_{\rm phot}$ measurements to measure the surface density at the fifth-nearest neighbor, $\Sigma_5(z)$ (Figure 5). Using emcee (D. Foreman-Mackey et al. 2013) to fit $\theta_{\rm E}$ based on image separations (Figure 6), we determine the Einstein mass, $M(<\theta_{\rm E})$, and classify the deflector scale. Our main results are as follows:

1. Most strong lenses in the AGEL sample with galaxy- or group-scale Einstein masses ($M(<\theta_{\rm E}) < 13$) are embedded in massive group or cluster halos (Figure 10). Including information on (higher density) environments leads to significant qualitative improvements in convergence and source positions for lens modeling.
2. Einstein mass is not strongly correlated with halo mass for the 89 deflectors in our sample (Figure 10). Using $\Sigma_5(z)$ to quantify deflector environment (halo mass), we find 21 galaxy-, 26 group-, and 42 cluster-scale deflector environments (Table 1). In contrast, the classification of the deflector scale (Einstein mass) with $M(<\theta_{\rm E})$ finds 22 galaxy-, 59 group-, and 8 cluster-scale deflectors. We consider several factors—including biases in the original AGEL selection and lensing by satellites in clusters—that may lead to the observed discrepancy between the two deflector properties.

3. The deflectors in AGEL systems are embedded in denser fields than control fields (Figure 12). Our result is inconsistent with past comparisons of lens environments with control fields. We hypothesize that the higher $\theta_{\rm E}$ of the AGEL distribution biases the AGEL fields to denser group and cluster environments. We hypothesize that more massive halos are more efficient at lensing, and that the AGEL survey selection favors brighter, more highly magnified source images.
4. We provide to the community detailed measurements of the 89 AGEL systems that include EAZY's $z_{\rm phot}$ and DECaLS' $r$-mag measurements to, e.g., quantify environment for lens modeling (see Tables 2, 3, and 4 in Appendix B). We also provide parameters including Einstein radii and masses, spectroscopic information, and our deflector environment measurements.

Our results are a cautionary tale that simplifying assumptions of the deflector scale do not account for the larger-scale properties of deflector host halos. Our results also suggest that more work is needed to address selection biases in current and upcoming lens searches. As we discover $\mathcal{O}(10^5)$ lenses in the coming decade, we could substantially improve cosmological and galaxy evolution models using lensing. However, to realize the full potential of the lensing breakthrough, we must carefully consider how our measurements are impacted by systematics related to lensing environment and selection biases in our lens searches.

## Acknowledgments

This research was made possible by the Harvard College Research Program (funding received for Fall 2024, Spring 2025, and Fall 2025 semesters), the Harvard Program for Research in Science and Engineering (PRISE; Summer 2024), and Harvard's Summer Program for Undergraduates in Data Science (SPUDS; Summer 2024). Parts of this research were conducted by the Australian Research Council Centre of Excellence for All Sky Astrophysics in 3 Dimensions (ASTRO 3D), through project No. CE170100013. K.G. and T.B. acknowledge support from Australian Research Council grant DP230101775. T.J. gratefully acknowledges support from the National Science Foundation through grant AST-2108515, the Gordon and Betty Moore Foundation through grant GBMF8549, NASA through grant HST-GO-16773, and a UC Davis Chancellor's Fellowship. A.H. acknowledges support from the ERC grant FIRSTLIGHT, Slovenian national research agency ARIS through grants N1-0238 and P1-0188. The authors would like to thank Keerthi Vasan G.C. for his valuable comments, and Benjamin D. Johnson for his advice on SED fitting.

The authors thank the referee for the constructive report. Some of the data used in this work were obtained at the W. M. Keck Observatory, which is operated as a scientific partnership among the California Institute of Technology, the University of California, and the National Aeronautics and Space Administration. The Observatory was made possible by the generous financial support of the W. M. Keck Foundation. The authors wish to recognize and acknowledge the very significant cultural role and reverence that the summit of Maunakea has always had within the indigenous Hawai'ian community. We are most fortunate to have the opportunity to conduct observations from this mountain.

This work was carried out at Harvard University, which is located on the traditional and ancestral land of the Massachusett, the original inhabitants of what is now Boston and Cambridge. We pay respect to the people of the Massachusett Tribe, past and present, and honor the land itself, which remains sacred to the Massachusett People.

## Data Availability

Deflector environment data for all of our systems are available at the Harvard Dataverse: doi: 10.7910/DVN/GYDQQP. For each system, we provide (1) a .csv file with $r$-band magnitudes, sky coordinates, and redshifts for all objects with fitted photo-$z$; (2) RGB LOS cutouts constructed from the DECaLS $g$, $r$, and $z$ bands, with fitted redshifts overlaid for all objects with measured photo-$z$; and (3) RGB LOS cutouts showing $r$-band magnitudes for all objects with measured photo-$z$.

*Facilities*: DECam (DECaLS), HST (WFC3), Keck (DEMOS), Keck (ESI), Keck (KCWI), Keck (MOSFIRE), Keck (LRIS), Keck (NIRES), VLT (X-SHOOTER), SDSS-BOSS.

*Software*: Astropy (Astropy Collaboration et al. 2013, 2018, 2022), DS 9 (W. A. Joye & E. Mandel 2003), EAZY (G. B. Brammer et al. 2008), emcee (D. Foreman-Mackey et al. 2013), GLEE (S. H. Suyu & A. Halkola 2010), lenstronomy (J. D. Hunter 2007; S. Birrer & A. Amara 2018), NumPy (C. R. Harris et al. 2020), TOPCAT (M. B. Taylor 2005).

## Appendix A
## Interesting Phenomena

In Tables 2, 3, and 4, the column "characteristics" lists systems we highlight for additional investigation. We henceforth refer to all three tables as the "classification table." We include four areas of interest based on deflector scale and environment measurements: (1) galaxy-scale lenses in cluster candidates, (2) galaxy-scale lenses in group candidates, (3) group-scale lenses, and (4) compound lenses.

### *A.1. Galaxy-scale Lenses in Cluster Candidates*

We label galaxy-scale deflectors in cluster environments as "GaGC" in our classification table. We detect seven such systems. Such lensing configurations enable probes of inner halo density profiles, which are useful for testing and refining $\Lambda$CDM (e.g., A. Ragagnin et al. 2022; G. Angora et al. 2023; C. Cerny et al. 2025). This sample may also be useful for explaining the discrepancy between lensing efficiency in $\Lambda$CDM-predicted clusters versus observed clusters (M. Meneghetti et al. 2020).

### *A.2. Galaxy-scale Lenses in Galaxy Groups*

We label galaxy-scale lenses in group environments as "GaGG" in our classification tables. As galaxy groups complete the continuum from galaxies to clusters, they are a useful probe of large-scale structure evolution. Galaxy groups are also more diverse than their cluster counterparts (M. Limousin et al. 2009). "GaGG" systems enable measurements of the galaxy–halo connection (e.g., M. Limousin et al. 2009; G. Zacharegkas et al. 2021), the concentration parameter $c_{200}$, and density profiles of dark matter halos (A. B. Newman et al. 2015). They may also help fix the concentration–mass and mass–luminosity relations across the entire mass spectrum (A. More et al. 2012).

We identify seven galaxy-scale deflectors in group-scale environments. Our sample provides a large sample of such lenses for a future analysis.

### *A.3. Group-scale Lenses*

We label group-scale deflector candidates as "GrSL" in our classification tables. We find 59 lenses with Einstein masses at the scale of groups in our sample. We expect to observe many more group-scale lenses in future surveys. Seven of the eight group-scale lenses are in cluster-scale deflector environments.

Although they are not located in groups on their own, group-scale lenses enable larger-scale probes of a cluster of dark matter halos. Group-scale lenses also possess simpler mass distributions, which could improve constraints on dark matter halo profiles and cosmological parameters (C. Lemon et al. 2024).

### *A.4. Compound Lenses in Galaxy or Group Environments*

Compound lenses, also referred to as double source plane lenses, occur when multiple sources at different redshifts are lensed by the same foreground deflector, with the nearer source(s) also contributing to the lensing of more distant source(s) (T. E. Collett & D. J. Bacon 2015). We label compound lenses as "CMPD" in our classification tables. We identify five compound lenses based on our spectroscopic data, but preliminary modeling of our galaxy-scale lens sample in L. Iwamoto et al. (2026, in preparation) suggests that the AGEL sample may contain several other compound lens candidates without spectroscopic confirmation.

Compound lenses enable independent measurements of cosmological parameters such as the Hubble constant, $H_0$, and the dark energy equation of state parameter, $w$ (T. E. Collett & M. W. Auger 2014). Two AGEL survey analyses with compound lenses in N. Sahu et al. (2025) and D. J. Bowden et al. (2025) show that increasing compound lens sample sizes substantially improves constraints on cosmological parameters. N. Sahu et al. (2025) include an analysis of compound lenses of AGEL1507, a galaxy-scale deflector in a galaxy environment; D. J. Bowden et al. (2025) also use AGEL0353, a galaxy-scale deflector in a group environment.

In our classification of compound lenses, we only choose deflectors that (1) are visually classified as galaxy-scale deflectors and (2) are located in galaxy or group environments. Four of the systems have group-scale Einstein mass classifications, but a visual inspection of DECaLS photometry shows only one central deflector. Our sparse environment requirement minimizes the impact of environment on the lensing profile, which will improve constraints on cosmological parameters.

## Appendix B
## Full Classification Tables

The following tables give our measured deflector scale and environment parameters for the 89 systems measured in our paper. Tables 2, 3, and 4 show parameters for systems in galaxy, group, and cluster environments, respectively.

**Table 2**
Full Table of Measurements and Classifications for Lenses with Galaxy-scale Deflector Environment Classifications

| AGEL ID (1) | $z_{spec,defl}$ (2) | $z_{spec,src}$ (3) | $\theta_E$ (arcsec) (4) | $M(<\theta_E)$ (5) | $\Sigma_5$ ($z$) (6) | $N_{LOS}$ (7) | $N_{env}$ (8) | Defl. Scale (9) | Defl. Env. (10) | Vis. Defl. (11) | Characteristics (12) |
|---|---|---|---|---|---|---|---|---|---|---|---|
| AGEL213758-012924 | 0.274 | 1.458 | 2.1 ± 0.9 | 11.8 ± 0.3 | 6.5 | 49 | 1 | Galaxy | Galaxy | Galaxy | ⋯ |
| AGEL233610-020735 | 0.494 | 2.662 | 1.7 ± 0.8 | 11.8 ± 0.3 | 30.4 | 27 | 4 | Galaxy | Galaxy | Galaxy | ⋯ |
| AGEL224621+223338 | 0.531 | 2.259 | 1.8 ± 0.9 | 11.9 ± 0.3 | 15.6 | 44 | 2 | Galaxy | Galaxy | Galaxy | ⋯ |
| AGEL004257-371858 | 0.879 | 3.094 | 1.4 ± 0.8 | 11.9 ± 0.4 | 38.3 | 17 | 4 | Galaxy | Galaxy | Galaxy | ⋯ |
| AGEL053724-464702 | 0.352 | 2.344 | 2.4 ± 0.8 | 12.0 ± 0.3 | 13.8 | 30 | 2 | Galaxy | Galaxy | Galaxy | ⋯ |
| AGEL231935+115016 | 0.541 | 1.99 | 2.2 ± 0.9 | 12.1 ± 0.3 | 15.7 | 16 | 2 | Group | Galaxy | Galaxy | GrSL |
| AGEL092315+182943 | 0.873 | 2.417 | 1.7 ± 1.0 | 12.1 ± 0.4 | 38.2 | 15 | 4 | Group | Galaxy | Galaxy | GrSL |
| AGEL142719-064515 | 0.265 | 1.512 | 3.5 ± 1.0 | 12.2 ± 0.2 | 0.0 | 85 | 0 | Group | Galaxy | Galaxy | GrSL |
| AGEL014253-183116 | 0.637 | 2.47 | 2.2 ± 1.0 | 12.2 ± 0.3 | 33.3 | 17 | 4 | Group | Galaxy | Galaxy | GrSL |
| AGEL025052-552412 | 0.718 | 2.478 | 2.1 ± 0.8 | 12.2 ± 0.3 | 26.2 | 29 | 3 | Group | Galaxy | Galaxy | GrSL |
| AGEL024303-000600 | 0.367 | 1.727[a], 0.506 | 3.0 ± 1.1 | 12.2 ± 0.3 | 0.0 | 42 | 0 | Group | Galaxy | Galaxy | GrSL, CMPD |
| AGEL222609+004142 | 0.647 | 1.896 | 2.2 ± 0.8 | 12.2 ± 0.3 | 33.6 | 25 | 4 | Group | Galaxy | N/A | GrSL |
| AGEL104041+185052 | 0.314 | 0.878 | 3.2 ± 0.8 | 12.3 ± 0.2 | 20.1 | 46 | 3 | Group | Galaxy | Galaxy | GrSL |
| AGEL000316-334804 | 0.656 | 1.834 | 2.5 ± 1.0 | 12.3 ± 0.3 | 33.7 | 34 | 4 | Group | Galaxy | Galaxy | GrSL |
| AGEL150745+052256 | 0.594 | 2.163[a], 2.6 | 3.5 ± 1.4 | 12.5 ± 0.3 | 24.4 | 25 | 3 | Group | Galaxy | Galaxy | GrSL, CMPD |
| AGEL104056-010359 | 0.25 | 1.21 | 6.7 ± 1.7 | 12.8 ± 0.2 | 19.1 | 79 | 3 | Group | Galaxy | Galaxy | GrSL |
| AGEL215122+134718 | 0.21 | 0.89 | 9.3 ± 2.1 | 13.0 ± 0.2 | 12.3 | 106 | 2 | Cluster | Galaxy | N/A | ⋯ |
| AGEL101807-000812 | 0.372 | 1.74 | N/A | N/A | 14.0 | 46 | 2 | Galaxy | Galaxy | Galaxy | ⋯ |
| AGEL023211+001339 | 0.892 | 2.365 | N/A | N/A | 38.5 | 30 | 4 | Group | Galaxy | N/A | GrSL |
| AGEL144149+144121 | 0.741 | 2.341 | N/A | N/A | 35.5 | 30 | 4 | Galaxy | Galaxy | Galaxy | ⋯ |
| AGEL110245+121111 | 0.925 | 2.806 | N/A | N/A | 39.2 | 14 | 4 | Galaxy | Galaxy | Galaxy | |

**Notes.** Rows are arranged in order of $M(<\theta_E)$. Tables 3 and 4 show the same measurements for group- and cluster-scale deflector environments, respectively. Columns: (1) is the AGEL ID as assigned in T. M. Barone et al. (2026), (2) and (3) are the spectroscopic deflector and source redshifts from T. M. Barone et al. (2026), (4) is our MCMC-measured $\theta_E$ ($''$); (5) is the mass enclosed by the measured $\theta_E$; (6) is the fifth-nearest neighbor measurement, $\Sigma_5(z)$, weighted by $z$; (7) is the number of objects along the line of sight; (8) is the number of deflector environment candidates; (9) is the deflector scale classification; (10) is the deflector environment classification; (11) classifies lenses as "Galaxy" if $\theta_E$ visually encloses one deflector in DECaLS imaging; and (12) is whether the lens is an interesting case described in Section Appendix A.

[a] If there are multiple sources, the source used to calculate $\theta_E$ and $M(<\theta_E)$. †: special characteristics of the lens (if applicable) as described in Appendix A. Galaxy-scale deflectors in cluster environments (GaGC; Section A.1), group-scale deflectors (GrSL; Section A.2), galaxy-scale deflectors in groups (GrSL; Section A.3), and compound galaxy-scale lenses in galaxy or group environments (CMPD; Section A.4).

**Table 3**
Group-scale Deflector Environment Catalog and Quantities

| AGEL ID (1) | $z_{\rm spec,defl}$ (2) | $z_{\rm spec,src}$ (3) | $\theta_{\rm E}$ (arcsec) (4) | $M(<\theta_{\rm E})$ (5) | $\Sigma_5$ ($z$) (6) | $N_{\rm LOS}$ (7) | $N_{\rm env}$ (8) | Defl. Scale (9) | Defl. Env. (10) | Vis. Defl. Scale (11) | Characteristics (12) |
|---|---|---|---|---|---|---|---|---|---|---|---|
| AGEL230522-000212 | 0.492 | 1.837 | 1.7 ± 0.9 | 11.8 ± 0.3 | 45.8 | 28 | 4 | Galaxy | Group | Galaxy | GaGG |
| AGEL085331+232155 | 0.307 | 2.188 | 2.1 ± 0.8 | 11.8 ± 0.3 | 42.6 | 35 | 5 | Galaxy | Group | Galaxy | GaGG |
| AGEL001310+004004 | 0.693 | 2.07 | 1.3 ± 0.7 | 11.8 ± 0.3 | 56.8 | 31 | 4 | Galaxy | Group | Galaxy | GaGG |
| AGEL014504-045551 | 0.635 | 1.96 | 1.6 ± 0.9 | 11.9 ± 0.4 | 49.7 | 37 | 5 | Galaxy | Group | Galaxy | GaGG |
| AGEL094328-015453 | 0.45 | 2.124 | 2.1 ± 0.9 | 12.0 ± 0.3 | 51.6 | 34 | 6 | Galaxy | Group | Galaxy | GaGG |
| AGEL091905+033639 | 0.444 | 2.195 | 2.1 ± 1.0 | 12.0 ± 0.3 | 80.2 | 36 | 11 | Galaxy | Group | Galaxy | GaGG |
| AGEL212252-005949 | 0.349 | 0.929 | 2.5 ± 1.2 | 12.1 ± 0.3 | 60.5 | 54 | 7 | Group | Group | Galaxy | GrSL |
| AGEL021225-085211 | 0.759 | 2.201 | 1.9 ± 1.0 | 12.1 ± 0.3 | 80.1 | 26 | 9 | Group | Group | Galaxy | GrSL |
| AGEL094412+322039 | 0.595 | 2.827 | 2.1 ± 0.9 | 12.1 ± 0.3 | 104.9 | 37 | 10 | Group | Group | Galaxy | GrSL |
| AGEL233459-640407 | 0.698 | 2.495 | 2.1 ± 0.9 | 12.1 ± 0.3 | 64.3 | 30 | 5 | Group | Group | Galaxy | GrSL |
| AGEL013639+000818 | 0.345 | 2.629 | 3.2 ± 0.9 | 12.2 ± 0.2 | 35.3 | 85 | 5 | Group | Group | N/A | GrSL |
| AGEL035346-170639 | 0.617 | 1.672[*], 1.46 | 2.1 ± 0.9 | 12.2 ± 0.3 | 63.7 | 31 | 6 | Group | Group | Galaxy | GrSL, CMPD |
| AGEL110154-060232 | 0.487 | 1.68 | 2.6 ± 1.0 | 12.2 ± 0.3 | 91.7 | 38 | 8 | Group | Group | Galaxy | GrSL |
| AGEL033717-315214 | 0.525 | 1.954 | 2.4 ± 1.3 | 12.2 ± 0.3 | 41.8 | 45 | 5 | Group | Group | Galaxy | GrSL |
| AGEL010238+015857 | 0.868 | 1.816 | 1.7 ± 0.7 | 12.2 ± 0.3 | 90.5 | 19 | 6 | Group | Group | Galaxy | GrSL |
| AGEL230305-511502 | 0.511 | 2.569 | 3.0 ± 0.9 | 12.3 ± 0.2 | 92.9 | 46 | 6 | Group | Group | Galaxy | GrSL |
| AGEL013355-643413 | 0.323 | 2.093 | 3.4 ± 1.2 | 12.3 ± 0.3 | 57.1 | 96 | 7 | Group | Group | Galaxy | GrSL |
| AGEL171922+244117 | 0.528 | 2.277 | 3.3 ± 1.2 | 12.4 ± 0.3 | 49.2 | 31 | 6 | Group | Group | N/A | GrSL |
| AGEL043806-322852 | 0.34 | 0.92 | 4.0 ± 1.1 | 12.5 ± 0.2 | 56.8 | 75 | 5 | Group | Group | N/A | GrSL |
| AGEL013442+043350 | 0.551 | 1.568, 2.03[*] | 3.3 ± 1.4 | 12.5 ± 0.3 | 111.7 | 39 | 9 | Group | Group | Galaxy | GrSL, CMPD |
| AGEL014556+040229 | 0.783 | 2.36 | 3.2 ± 0.8 | 12.6 ± 0.2 | 86.8 | 17 | 5 | Group | Group | N/A | GrSL |
| AGEL140839+253104 | 0.663 | 1.287 | 3.4 ± 0.9 | 12.7 ± 0.2 | 85.2 | 22 | 9 | Group | Group | Galaxy | GrSL |
| AGEL133041+044015 | 0.336 | 1.168 | 5.4 ± 1.4 | 12.7 ± 0.2 | 46.8 | 71 | 5 | Group | Group | N/A | GrSL |
| AGEL093333+091919 | 0.743 | 2.434[*], 2.435 | 4.0 ± 1.0 | 12.7 ± 0.2 | 93.4 | 36 | 8 | Group | Group | N/A | GrSL |
| AGEL212326+015312 | 0.591 | 1.18 | 4.5 ± 1.0 | 12.9 ± 0.2 | 60.8 | 28 | 8 | Group | Group | N/A | GrSL |
| AGEL080820+103142 | 0.475 | 1.452, 1.303[*] | N/A | N/A | 69.4 | 41 | 9 | Galaxy | Group | Galaxy | GaGG, CMPD |

**Note.** Column descriptors are the same as in Table 2.[*]

**Table 4**
Cluster-scale Deflector Environment Catalog and Quantities

| AGEL ID (1) | $z_{spec,defl}$ (2) | $z_{spec,src}$ (3) | $\theta_E$ (arcsec) (4) | $M(<\theta_E)$ (5) | $\Sigma_5$ ($z$) (6) | $N_{LOS}$ (7) | $N_{env}$ (8) | Defl. Scale (9) | Defl. Env. (10) | Vis. Defl. Scale (11) | Characteristics (12) |
| --- | --- | --- | --- | --- | --- | --- | --- | --- | --- | --- | --- |
| AGEL013719-083056 | 0.563 | 2.997 | 1.7 ± 0.7 | 11.8 ± 0.3 | 170.5 | 49 | 11 | Galaxy | Cluster | Galaxy | GaGC |
| AGEL000224-350716 | 0.573 | 1.549 | 1.7 ± 0.6 | 11.9 ± 0.3 | 255.5 | 33 | 8 | Galaxy | Cluster | Galaxy | GaGC |
| AGEL084943+294328 | 0.68 | 2.055 | 1.7 ± 0.7 | 12.0 ± 0.3 | 489.4 | 35 | 10 | Galaxy | Cluster | Galaxy | GaGC |
| AGEL014235-164818 | 0.619 | 2.309 | 1.9 ± 1.0 | 12.0 ± 0.3 | 562.7 | 35 | 16 | Galaxy | Cluster | Galaxy | GaGC |
| AGEL231112-454658 | 0.501 | 1.555 | 1.9 ± 0.9 | 12.0 ± 0.3 | 228.0 | 53 | 20 | Galaxy | Cluster | Galaxy | GaGC |
| AGEL003727-413150 | 0.708 | 3.259 | 1.8 ± 0.9 | 12.0 ± 0.3 | 197.1 | 29 | 13 | Galaxy | Cluster | Galaxy | GaGC |
| AGEL020613-011417 | 0.632 | 1.302 | 1.5 ± 1.1 | 12.0 ± 0.4 | 194.8 | 41 | 12 | Galaxy | Cluster | Galaxy | GaGC |
| AGEL001424+004145 | 0.569 | 1.368 | 2.1 ± 0.9 | 12.2 ± 0.3 | 280.5 | 14 | 5 | Group | Cluster | Galaxy | GrSL |
| AGEL101847-012132 | 0.388 | 1.43 | 2.7 ± 0.9 | 12.2 ± 0.3 | 355.3 | 48 | 9 | Group | Cluster | Galaxy | GrSL |
| AGEL000645-442950 | 0.495 | 2.097 | 2.6 ± 0.9 | 12.2 ± 0.3 | 279.1 | 37 | 11 | Group | Cluster | Galaxy | GrSL |
| AGEL022931-290816 | 0.853 | 1.679 | 1.7 ± 0.8 | 12.2 ± 0.3 | 192.6 | 25 | 5 | Group | Cluster | Galaxy | GrSL |
| AGEL211243+000921 | 0.445 | 2.359 | 3.2 ± 0.9 | 12.3 ± 0.2 | 195.4 | 38 | 9 | Group | Cluster | Galaxy | GrSL |
| AGEL211005-563931 | 0.497 | 1.182 | 2.4 ± 0.9 | 12.3 ± 0.3 | 203.0 | 40 | 9 | Group | Cluster | Galaxy | GrSL |
| AGEL010257-291122 | 0.273 | 0.815 | 3.6 ± 1.2 | 12.3 ± 0.3 | 448.3 | 134 | 19 | Group | Cluster | Galaxy | GrSL |
| AGEL221912-434835 | 0.71 | 2.167 | 2.4 ± 1.1 | 12.3 ± 0.3 | 139.7 | 30 | 11 | Group | Cluster | Galaxy | GrSL |
| AGEL155417+044339 | 0.78 | 1.72 | 2.1 ± 0.8 | 12.3 ± 0.3 | 413.0 | 26 | 9 | Group | Cluster | Galaxy | GrSL |
| AGEL211627-594702 | 0.395 | 1.411 | 3.4 ± 1.2 | 12.4 ± 0.3 | 118.8 | 53 | 10 | Group | Cluster | Galaxy | GrSL |
| AGEL233552-515218 | 0.566 | 2.224 | 3.2 ± 1.2 | 12.4 ± 0.3 | 194.5 | 43 | 16 | Group | Cluster | Galaxy | GrSL |
| AGEL153929+165016 | 0.41 | 1.253 | 3.4 ± 1.2 | 12.4 ± 0.3 | 191.3 | 74 | 21 | Group | Cluster | Galaxy | GrSL |
| AGEL214915-001252 | 0.452 | 1.943 | 3.9 ± 1.2 | 12.5 ± 0.2 | 228.7 | 43 | 13 | Group | Cluster | N/A | GrSL |
| AGEL091126+141757 | 0.546 | 1.206 | 3.0 ± 1.1 | 12.5 ± 0.3 | 665.2 | 43 | 16 | Group | Cluster | Galaxy | GrSL |
| AGEL004144-233905 | 0.439 | 2.203 | 4.5 ± 1.3 | 12.6 ± 0.2 | 532.9 | 67 | 15 | Group | Cluster | N/A | GrSL |
| AGEL201419-575701 | 0.713 | 2.191 | 3.5 ± 1.0 | 12.6 ± 0.2 | 304.6 | 34 | 14 | Group | Cluster | Galaxy | GrSL |
| AGEL014106-171324 | 0.608 | 2.436 | 3.9 ± 1.1 | 12.6 ± 0.2 | 161.3 | 30 | 5 | Group | Cluster | N/A | GrSL |
| AGEL002700-041324 | 0.495 | 1.46 | 3.6 ± 1.3 | 12.6 ± 0.3 | 313.4 | 68 | 21 | Group | Cluster | N/A | GrSL |
| AGEL212512-650427 | 0.779 | 2.221 | 3.1 ± 1.2 | 12.6 ± 0.3 | 135.0 | 24 | 6 | Group | Cluster | Galaxy | GrSL |
| AGEL162300+213721 | 0.761 | 1.727 | 3.4 ± 0.9 | 12.7 ± 0.2 | 141.3 | 30 | 8 | Group | Cluster | N/A | GrSL |
| AGEL004827+031117 | 0.357 | 2.37 | 5.2 ± 1.5 | 12.7 ± 0.2 | 243.2 | 82 | 10 | Group | Cluster | N/A | GrSL |
| AGEL011759-052718 | 0.58 | 2.065 | 4.8 ± 1.2 | 12.8 ± 0.2 | 262.2 | 28 | 10 | Group | Cluster | N/A | GrSL |
| AGEL091935+303156 | 0.427 | 1.811 | 5.4 ± 1.8 | 12.8 ± 0.3 | 116.0 | 50 | 10 | Group | Cluster | N/A | GrSL |
| AGEL224405+275916 | 0.357 | 0.959 | 5.3 ± 1.8 | 12.8 ± 0.3 | 253.3 | 64 | 14 | Group | Cluster | Galaxy | GrSL |
| AGEL103027-064109 | 0.468 | 1.581 | 5.4 ± 1.3 | 12.9 ± 0.2 | 139.0 | 60 | 9 | Group | Cluster | N/A | GrSL |
| AGEL132304+034319 | 0.353 | 1.016 | 6.6 ± 1.6 | 13.0 ± 0.2 | 266.1 | 80 | 13 | Group | Cluster | N/A | GrSL |
| AGEL042439-331742 | 0.564 | 1.188 | 5.0 ± 1.8 | 13.0 ± 0.3 | 334.1 | 52 | 17 | Group | Cluster | Galaxy | GrSL |
| AGEL152560+084639 | 0.602 | 1.95 | 6.4 ± 1.4 | 13.1 ± 0.2 | 838.8 | 48 | 14 | Cluster | Cluster | N/A | ⋯ |
| AGEL123809+150151 | 0.572 | 1.16 | 6.8 ± 1.5 | 13.3 ± 0.2 | 634.0 | 52 | 15 | Cluster | Cluster | N/A | ⋯ |
| AGEL235934+020824 | 0.429 | 1.12 | 9.5 ± 2.1 | 13.4 ± 0.2 | 438.8 | 83 | 18 | Cluster | Cluster | N/A | ⋯ |
| AGEL002527+101107 | 0.463 | 2.397 | 12.1 ± 2.5 | 13.5 ± 0.2 | 377.7 | 51 | 15 | Cluster | Cluster | N/A | ⋯ |
| AGEL204312-060954 | 0.793 | 1.89 | 8.8 ± 1.9 | 13.5 ± 0.2 | 170.6 | 21 | 9 | Cluster | Cluster | N/A | ⋯ |
| AGEL144133-005401 | 0.54 | 1.667 | 11.2 ± 1.9 | 13.6 ± 0.2 | 642.3 | 63 | 18 | Cluster | Cluster | N/A | ⋯ |
| AGEL060357-355806 | 0.488 | 0.962[a] | 17.1 ± 3.0 | 14.0 ± 0.2 | 260.4 | 85 | 22 | Cluster | Cluster | N/A | ⋯ |
| AGEL014327-085021 | 0.737 | 2.755 | N/A | N/A | 437.4 | 32 | 10 | Group | Cluster | N/A | GrSL |

**Notes.** Column descriptors are the same as in Table 2. * If there are multiple sources, the source used to calculate $\theta_E$ and $M(<\theta_E)$.
[a] AGEL060357-355806 has several sources in addition to the one we list (see W. Sheu et al. 2024).

## ORCID iDs

William J. Gottemoller https://orcid.org/0009-0008-5372-1318
Nandini Sahu https://orcid.org/0000-0003-0234-6585
Rodrigo Cordova Rosado https://orcid.org/0000-0002-7967-7676
Leena Iwamoto https://orcid.org/0009-0006-4812-2033
Courtney B. Watson https://orcid.org/0000-0001-8456-4142
Kim-Vy H. Tran https://orcid.org/0000-0001-9208-2143
A. Makai Baker https://orcid.org/0009-0000-6139-3280
Tania M. Barone https://orcid.org/0000-0002-2784-564X
Duncan J. Bowden https://orcid.org/0009-0008-6114-1401
Karl Glazebrook https://orcid.org/0000-0002-3254-9044
Anishya Harshan https://orcid.org/0000-0001-9414-6382
Tucker Jones https://orcid.org/0000-0001-5860-3419
Glenn G. Kacprzak https://orcid.org/0000-0003-1362-9302
Camryn M. Neches https://orcid.org/0009-0008-3060-459X